\documentclass[fleqn,usenatbib]{mnras}

\usepackage{newtxtext,newtxmath}

\usepackage[T1]{fontenc}
\usepackage{ae,aecompl}

\usepackage{graphicx}	
\usepackage{amsmath}	
\usepackage{amssymb}	
\usepackage{subfigure, color}
\usepackage{xcolor}
\usepackage{cleveref}

\title[Global trends in winds of M dwarf stars]{Global trends in winds of M dwarf stars}

\author[A. L. Mesquita and A. A. Vidotto]{
A. L. Mesquita,$^{1}$\thanks{E-mail: mesquita@tcd.ie}
and A. A. Vidotto$^{1}$
\\ 
$^{1}$School of Physics, Trinity College Dublin, The University of Dublin, Dublin 2, Ireland}

\date{Accepted XXX. Received YYY; in original form ZZZ}

\pubyear{2019}

\begin{document}
\label{firstpage}
\pagerange{\pageref{firstpage}--\pageref{lastpage}}
\maketitle

\begin{abstract}
M dwarf stars are currently the main targets in searches for potentially habitable planets. However, their winds have been suggested to be harmful to planetary atmospheres. Here, in order to better understand the winds of M dwarfs and also infer their physical properties, we perform a one-dimensional magnetohydrodynamic parametric study of winds of M dwarfs that are heated by dissipation of Alfv\'{e}n waves. These waves are triggered by sub-surface convective motions and propagate along magnetic field lines. Here, we vary the magnetic field strength $B_0$ and density $\rho_0$ at the wind base (chromosphere), while keeping the same relative wave amplitude ($0.1 B_0$) and dissipation lenghtscale. Our simulations thus range from low plasma-$\beta$ to high plasma-$\beta$ (0.005 to 3.7). We find that our winds very quickly reach isothermal temperatures with mass-loss rates $\dot{M} \propto \rho_0^2$. We compare our results with Parker wind models and find that, in the high-$\beta$ regime, both models agree. However, in the low-$\beta$ regime, the Parker wind underestimates the terminal velocity by around one order of magnitude and $\dot{M}$ by several orders of magnitude. We also find that M dwarfs could have chromospheres extending to 18\% to 180\% of the stellar radius. We apply our model to the planet-hosting star GJ 436 and find, from X-ray observational constraints, $\dot{M}<7.6\times 10^{-15}\,M_{\odot}~\text{yr}^{-1}$. This is in agreement with values derived from the Lyman-$\alpha$ transit of GJ 436b, indicating that spectroscopic planetary transits could be used as a way to study stellar wind properties.
\end{abstract}

\begin{keywords}
MHD -- stars: low-mass -- stars: winds, outflows -- stars: mass-loss -- planetary systems: GJ 436
\end{keywords}



\section{Introduction}
M dwarf stars are the most common type of stars in our Galaxy. They are small, main-sequence stars with masses smaller than $\sim 0.5 M_{\odot}$, low surface temperatures and low brightness. One of the most interesting features in M dwarfs is their close-in habitable zone, which is defined as the extended area away from the star where an orbiting planet could have liquid water on its surface \citep{Kasting1993, Selsis2007}. Because M dwarf stars have low luminosities, their habitable zones are much closer in, which makes it easier to observe exoplanets in their habitable zones due to current biases in planet detection. For a M dwarf with 0.5\,$M_{\odot}$, for example, the habitable zone is at $\sim$ 0.2 -- 0.4 astronomical units \citep{Kasting1993, Selsis2007}. However, one potential issue for planet habitability is that main-sequence M dwarfs remain active for a long fraction of their lives, generating strong kG magnetic fields \citep{Morin2010, Lang2014, See2019, Shulyak2019}. A star with strong activity can generate strong flares, winds and coronal mass ejections, which can affect the exoplanets orbiting their habitable zones as well as exoplanet habitability \citep{Khodachenko2007, Vida2017, Tilley2019}.

In addition to consequences to planetary habitability, stellar winds play an essential role in stellar evolution \citep{Matt2015, Johnstone2015}. However, M dwarfs, similar to other cool dwarf stars, have rarefied winds and, as a consequence, it is difficult to directly measure them \citep{Wood2004, Vidotto2017, Jardine2019}. There are some techniques developed to infer the mass-loss rates of cool dwarf stars, such as radio emission analysis \citep{Panagia1975, Lim1996, Fichtinger2017, VidottoD2017}, or the identification of X-ray emission generated due to the interaction between ionized wind particles with neutral atoms from the interstellar medium \citep{Wargelin2002, Jardine2019}. Another more successful method used to detect stellar winds is related to the study of stellar Ly-$\alpha$ line absorption when the stellar wind exchange charges with a neutral or partially neutral interstellar medium \citep{Wood2002, Wood2005}. By studying Ly-$\alpha$ observations of the binary system $\alpha$ Centauri (G2 + K0) and its distant companion star Proxima Centauri (M5.5), \citet{Wood2001} predicted a mass-loss rate upper limit of $\dot{M}<4\times 10^{-15}\,M_{\odot}~\text{yr}^{-1}$ for Proxima Centauri. A recently proposed method is to use the exoplanet atmosphere interaction with the host star wind to infer some properties of the local stellar wind \citep{Vidotto2017}. These techniques have provided some constraints on the winds of M dwarfs, but still a full picture does not yet exist.

In the present work, we turn to numerical simulations to investigate stellar winds of M dwarfs. There are still not many numerical studies dedicated to the winds of M dwarfs \citep[e.g.][]{Vidotto2014, Garraffo2016, Vidotto2017}. In cool dwarfs, it is common to study winds by adopting a Parker wind model \citep{Parker1958}, which consists of a stellar wind with constant temperature. One weakness of isothermal winds is that, by assuming that the temperature is constant, we cannot derive the detailed structure of the wind energetics, such as, heating and cooling. In our work, we assume that the winds of M dwarfs are heated by magnetic processes, similar to the solar wind. For that, we use a model that considers the presence of Alfv\'{e}n-waves to drive the winds of M dwarfs. With this, we can better investigate the physical processes of the wind acceleration mechanism and of its heating.

Alfv\'{e}n waves are magnetohydrodynamic (MHD) waves that propagate with an Alfv\'{e}n velocity $v_{A}=B/\sqrt{4\pi\rho}$, where $B$ is the magnetic field and $\rho$ is the density. In 1942, \citeauthor{Alfven1942} hypothesized that MHD waves generated in the lower layers of the Sun could be associated with  sunspots. Later on, \citet{Schatzman1949} proposed that MHD waves were responsible by the coronal heating. Alfv\'{e}n waves are still one of the hypothesis to explain the temperature gradient in the Sun's atmosphere (e.g., \citealt{Winebarger2004, DeMoortel2015}). Alfv\'{e}n waves are generated if oscillations are induced at the magnetic field at the base of the wind. The dissipation of energy and momentum associated with the wave propagation can lead to the acceleration of the outer atmosphere in the form of an Alfv\'{e}n-wave driven wind \citep{Hartmann1980, Vidotto2006}.

In this paper, we perform a parametric study of winds of M dwarf stars, using an Alfv\'{e}n-wave driven stellar wind model to understand the winds of M dwarfs and also infer their properties, like mass-loss rates, velocities, etc. This paper consists of the following sections. In \Cref{sec:model}, we describe our stellar wind model and the simulation parameters. Our results for the wind structure and general trends of M dwarfs are presented in \Cref{sec:wind-total}, followed by a discussion about the chromospheric size of M dwarfs and an application to the planet-hosting star GJ 436 in \Cref{sec:applic}. Finally, we present a comparison with a Parker wind model and a discussion about the effects of the free input parameters in our simulation in \Cref{sec:discu} followed by conclusions in \Cref{sec:conc}.

\section{Alfv\'{e}n-wave driven stellar wind model}
\label{sec:model}
We perform one dimensional magnetohydrodynamic simulations to heat and drive the wind of M dwarf stars. Alfv\'{e}n waves are generated by perturbations induced in the magnetic field at the base of the wind. The waves accelerate the stellar atmosphere in the form of an Alfv\'{e}n-wave driven wind \citep{Hartmann1980, Holzer1983, MacGregor1994}. The model used in this work is based on \citet{Vidotto2010}, and we describe it next.
 
We numerically solve the time-independent MHD equations including momentum and energy equations:
\begin{equation}
    u\frac{du}{dr}=-\frac{GM_{\star}}{r^2}-\frac{1}{\rho}\frac{dP}{dr}-\frac{1}{2\rho}\frac{d\epsilon}{dr},
    \label{eq:momentum}
\end{equation}
\begin{equation}
    \rho u\frac{d}{dr}\left(\frac{u^2}{2}+\frac{5}{2}\frac{k_{B}T}{m}-\frac{GM_{\star}}{r}\right)+\rho u \frac{d}{dr}\left(\frac{F_{c}}{\rho u}\right)+\frac{u}{2}\frac{d\epsilon}{dr}=Q-P_{r},
    \label{eq:energy}
\end{equation}
where $u$ is the wind velocity, $r$ the radial coordinate, $G$ the gravitational constant, $M_{\star}$ the stellar mass, $P=\rho k_B T/m$ the gas pressure, $m$ the average mass of the wind particles, $\rho$ the wind density, $T$ the wind temperature, $\epsilon$ the energy density of the Alfv\'{e}n waves, $F_c$ the termal conduction, $Q$ the volumetric heating rate and $P_r$ is the volumetric radiative cooling rate.

The terms on the right-hand side of \Cref{eq:momentum} are the gravitational force, the gradient of the thermal pressure and the gradient of the wave pressure, respectively. The first, second and third terms on the left-hand side of \Cref{eq:energy} are related to the wind energy (kinetic energy, enthalpy and gravitational energy), the thermal conductivity and the rate at which the waves do work, respectively. The terms on the right-hand side of \Cref{eq:energy} are related to the wave heating and the radiative cooling. 

The energy density of the Alfv\'{e}n waves \citep{Hartmann1980} are given by:
\begin{equation}
    \epsilon=\epsilon_0\frac{M_0}{M}\left(\frac{1+M_0}{1+M}\right)\exp{\left[-\int_{r_0}^{r}\frac{1}{L}dr\right]},
\label{eq:density}
\end{equation}
where $M$ is the Mach number and $L$ is the damping length. In this paper, whenever we use the subscript ``0'', it represents a quantity calculated at the base of the wind, thus, in \Cref{eq:density}, $M_0$ is the Mach number at the wind base at $r=r_0$. Here, we assume the nonlinear damping mechanism for the waves, as this has been used in some solar wind models \citep{Suzuki2005, Suzuki2013}. We parametrise the non-linear damping mechanism, following the work of \citet{Jatenco1989}, by
\begin{equation}
    L=L_{0}\left(\frac{v_{A}}{v_{A0}}\right)^{4}\frac{\langle \delta v_{0}^{2}\rangle}{\langle \delta v^{2}\rangle}(1+M),
\end{equation}
with an initial length of $10\%$ of stellar radius ($L_0=0.1r_0$). Here, $\langle \delta v^{2}\rangle$ is the mean quadratic amplitude of the fluctuations in the wave velocity. The amplitude of the velocity fluctuations are connected with the amplitude of magnetic field fluctuations by energy equipartition
\begin{equation}
   \frac12 \rho \langle \delta v^{2}\rangle=\frac{\langle \delta B^{2}\rangle}{8 \pi}.
\end{equation}

Finally, the energy density of the wave is related to its flux as
\begin{equation}
    \phi_{A}=\epsilon v_{A}\left(1+\frac{3}{2}M\right).
\end{equation}

The thermal conduction flux is 
\begin{equation}
   F_c=-\kappa T^{5/2}\frac{dT}{dr}, 
\end{equation}
where $\kappa=10^{-6}\,\text{erg~cm}^{-1}\text{s}^{-1}\text{K}^{-1}$ is the Spitzer conductivity. The volumetric heating rate caused by wave dissipation is 
\begin{equation}
    Q=\frac{\epsilon}{L}(u+v_{A}),
\end{equation}
and the volumetric radiative cooling rate is
\begin{equation}
    P_{r}=\Lambda n_{e}n_{H},
\end{equation}
where $\Lambda$ is the cooling function which depends on the metallicity, $n_e$ is the electron density and $n_H$ is the total hydrogen density. In our simulations, we adopt the cooling function from \citet{Schure2009} for solar-like metallicity. Given the high temperatures our winds achieve, our winds are fully ionised through the simulation domain, which implies that $n_H=n_p=n_e$, where $n_p$ is the proton density.

We also numerically solve the mass conservation equation, assuming steady state
\begin{equation}
    \frac{d}{dr}\left(\rho u r^2\right) = 0.
    \label{eq:mass-con}
\end{equation}

We initially perform 134 simulations assuming spherical symmetry with the input parameters presented in \Cref{tab:input}. We use the values of mass and radius for an early M dwarf, similar to GJ 436. We adopt an open magnetic field line configuration with magnetic field oscillations induced at the base of the chromosphere. The initial perturbations in the magnetic field lines were set to be $10\%$ of magnetic field, $\sqrt{\langle\delta B_0^2\rangle}=0.1B_0$. Given our values of input magnetic fields our simulations are more appropriate for an inactive to moderately active star. Our simulations results in wave fluxes ranging from $7.9\times 10^{2}$ to $1.64\times 10^{6}$\,erg~cm$^{-2}$s$^{-1}$ at the base of the chromosphere.

\begin{table*}
 \caption{The top part of the table shows the input parameters of our simulations,
assumed at the base of the wind (chromosphere). The parameters at the bottom, below the line, are derived from
the input parameters.}
 \label{tab:input}
 \begin{tabular}{lccc}
  \hline
  Physical parameter & Symbol & Value & Unit\\
  \hline
  Stellar mass & $M_{\star}$ & 0.452 & $M_{\odot}$\\
  Stellar radius & $r_0$ & 0.437 & $R_{\odot}$\\
  Temperature & $T_0$ & $2\times 10^{4}$ & K \\
  Magnetic field & $B_0$ & 1 -- 10 & G\\
  Density & $\rho_0$ & (1 -- 90)$\times 10^{-15}$ & g~cm$^{-3}$\\
  Magnetic field perturbation & $\sqrt{\langle\delta B_0^2\rangle}$ & 0.1 & $B_0$\\
  Damping length & $L_0$ & 0.1 & $r_0$\\
  \hline
  Wave amplitude & $\sqrt{\langle\delta v_0^2\rangle}$ & 0.9 -- 25.4 & km~s$^{-1}$\\
  Wave flux & $\phi_{A0}$ & $7.9\times 10^{2}$ -- $1.64\times 10^{6}$ & erg~cm$^{-2}$s$^{-1}$\\
  Wave energy density & $\epsilon_{0}$ & (7.9 -- 790)$\times 10^{-4}$ & erg~cm$^{-3}$\\
  plasma $\beta$ at base & $\beta$ & 0.005 -- 3.7  \\
  \hline
 \end{tabular}
\end{table*}

To solve the set of coupled differential equations, we use a shooting method, in which the only physical solution is the one that passes through the Alfv\'{e}n point \citep[e.g.][]{Vidotto2006}. The Alfv\'{e}n point is the point where the wind velocity is equal to Alfv\'{e}n velocity, i.e., the distance where the Mach number is unit ($M=u/v_A=1$). This is an important parameter for calculating the angular momentum-loss rate, which we will discuss in \Cref{sec:wind-trends} \citep{Weber1967, Kraft1967}.

\section{Parametric study of winds of M-dwarfs}
\label{sec:wind-total}
\subsection{The structure of the wind} 
\label{sec:wind}
\begin{figure}
    \includegraphics[width=\columnwidth]{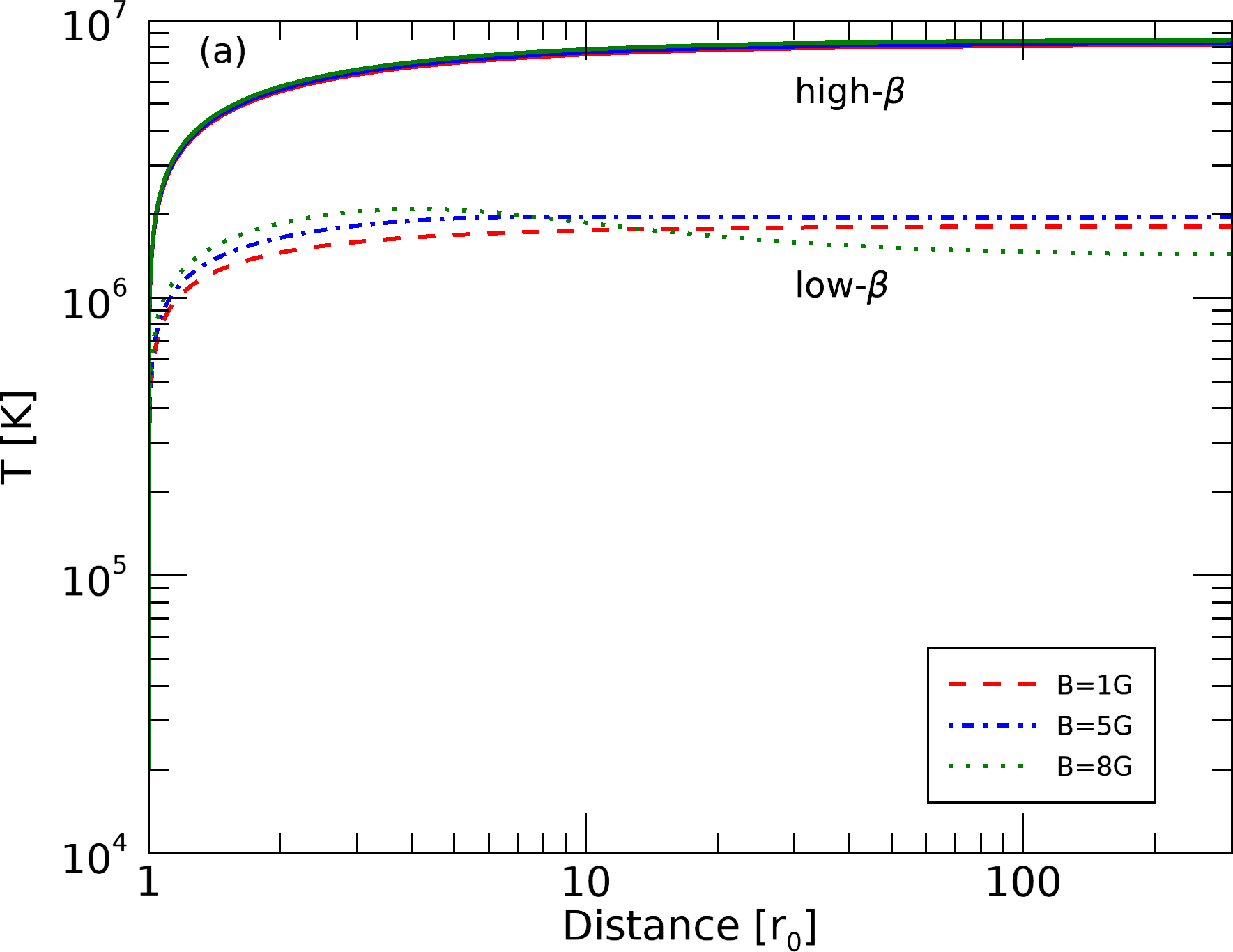}
    \includegraphics[width=\columnwidth]{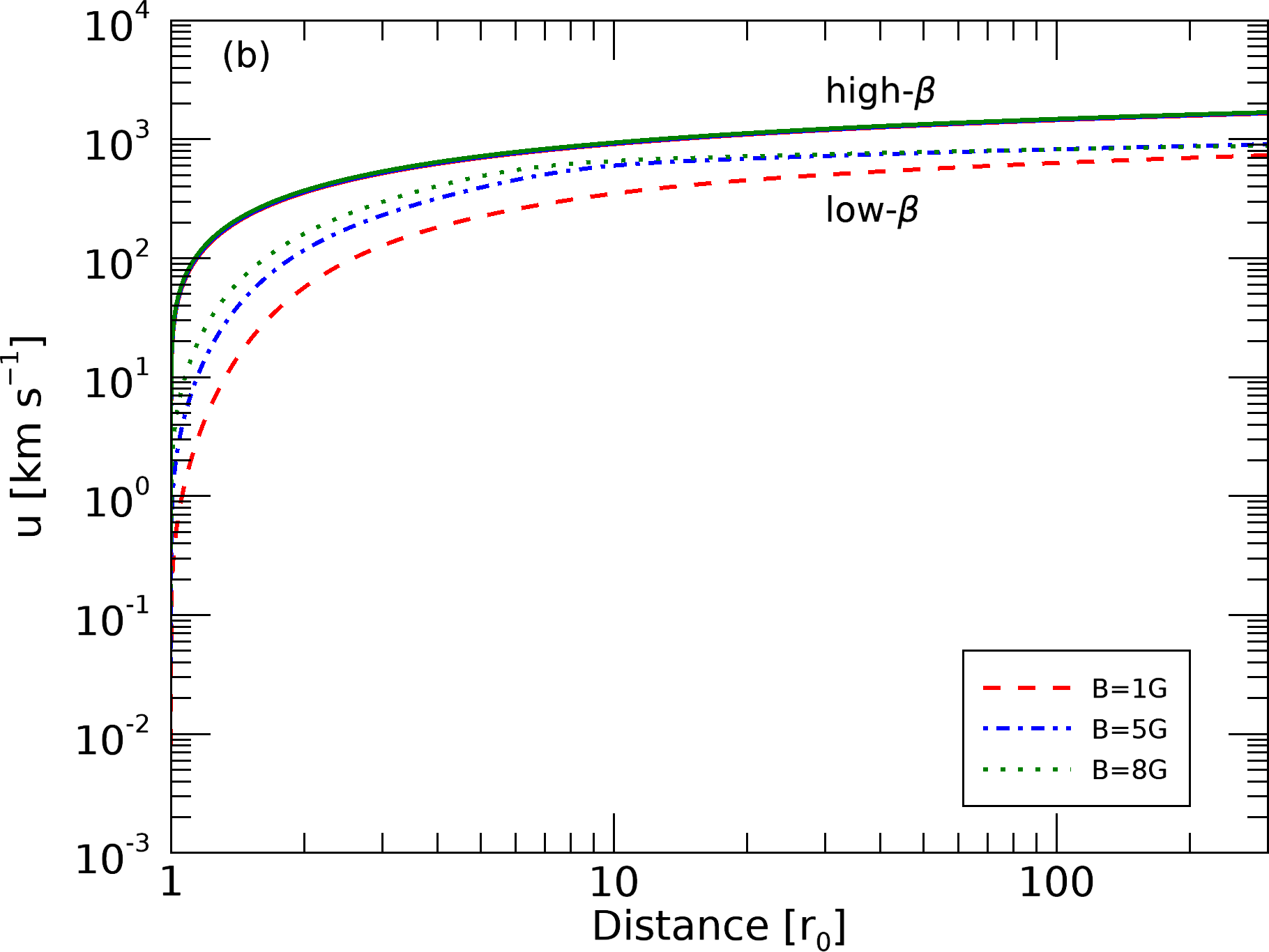}
    \includegraphics[width=\columnwidth]{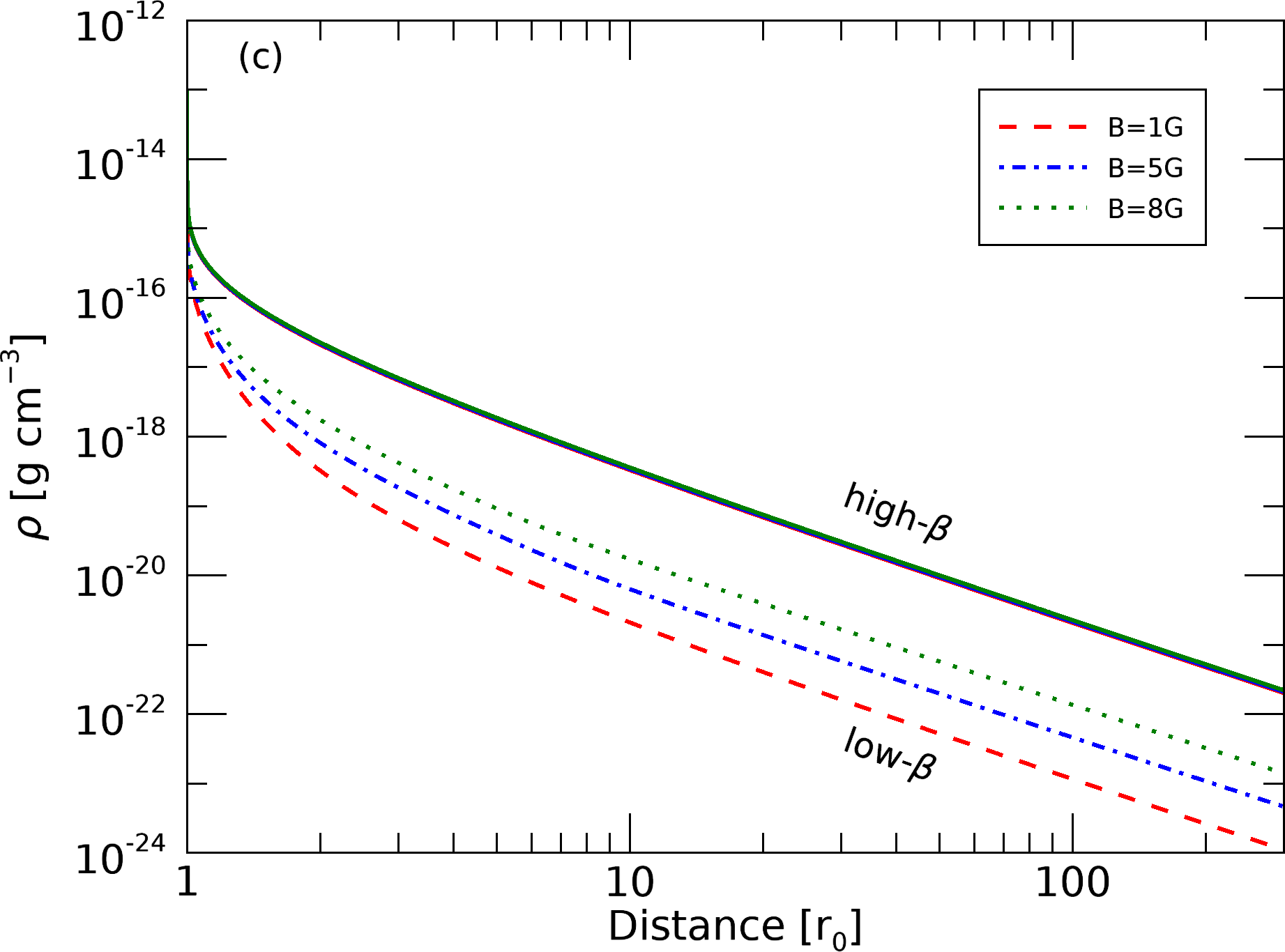}
  \caption{Temperature, velocity and density profiles for high-$\beta$ (solid lines) and low-$\beta$ (non-solid lines) ranges and magnetic fields 1 (red), 5 (blue) and 8\,G (green). The profiles for high-$\beta$ do not depend significantly on the magnetic field intensity. This is because these winds are thermally dominated. We use $\rho_0=9\times 10^{-15}\,\text{g~cm}^{-3}$ for low-$\beta$ cases and $\rho_0=9\times 10^{-14}\,\text{g~cm}^{-3}$ for high-$\beta$ cases.}
  \label{fig:profiles}
\end{figure}

To understand the wind properties, we analyze how temperature, velocity and density profiles are affected by different input parameters. \Cref{fig:profiles} shows wind profiles for different magnetic field intensities and for two ranges of base density. We separate the base densities in two ranges that we label as `low-$\beta$' for $\rho_0=(1-9)\times 10^{-15}\,\text{g~cm}^{-3}$ and `high-$\beta$' for $\rho_0=(1-9)\times 10^{-14}\,\text{g~cm}^{-3}$. The plasma $\beta$ parameter gives information about the balance between the gas pressure and the magnetic pressure and is given by:
\begin{equation}
   \beta=\frac{P}{P_{\text{mag}}}=\frac{8\pi\rho k_{B}T}{mB^{2}},
   \label{eq:beta}
\end{equation}
where $P_{\text{mag}}=B^{2}/8\pi$ is the magnetic pressure.

We see an overall higher wind temperature (\Cref{fig:profiles}-a) for higher base densities (high-$\beta$), and, to a lesser extent, higher temperatures are also seen with higher base magnetic fields. However, the magnetic field does not affect significantly the temperature profiles for high-$\beta$, which is seen in the similarities of all distance-profiles. The temperature profile displays a sudden rise before $\sim 1.5\,r_0$ and then reaches a flat profile. This flat profile is caused by conduction -- models of red supergiant winds without conductive fluxes, for example, do not show this \citep{Vidotto2006}. The plateau profile can be interpreted as M dwarfs having nearly isothermal winds. We will come back to this in \Cref{sec:parker}, when we compare our results with Parker winds, and the trends with plasma $\beta$ are discussed in \Cref{sec:wind-trends}

The overall velocity profiles (\Cref{fig:profiles}-b) are higher for high magnetic field intensities. These profiles show two different behaviours with $\beta$ parameter. For all our wind models, the terminal velocities vary from around $710$ to $3100\,\text{km~s}^{-1}$. The wind is rapidly accelerated by the transfer of momentum from the waves to the plasma (a consequence of the third term in \Cref{eq:momentum}) and then reaches an asymptotic profile. The acceleration process happens closer to the star for high-$\beta$ cases. For low-$\beta$, the velocity decreases with base density and for high-$\beta$ the velocity increases with base density. The velocity profile for high-$\beta$ is not significantly affected by the magnetic field intensity. 

The overall density profiles (\Cref{fig:profiles}-c) are higher for higher magnetic field intensities and base densities. Further away from the star, the wind is orders of magnitude less dense than at the base, which demonstrates that winds of M dwarfs can be very rarefied, like the solar wind. Similarly to the other profiles studied here, the density profile for high-$\beta$ is not significantly affect by the intensity of the base magnetic field. At large distances, the density profile falls with $r^2$ as a consequence of mass conservation (\Cref{eq:mass-con}) and the asymptotic wind speed. At small distances, the nearly exponential decrease in density is due to the rapid increase observed in the velocity profile.

All profiles show a very clear trend according to base density range. The physical explanation for it lies on the $\beta$ parameter. Beta smaller than one ($\beta<1$) indicates that magnetic field plays a major role in the wind and beta greater than one ($\beta>1$) indicates that the thermal forces dominate. In our simulations, we do not change the temperature at the base, which means that the only parameters influencing $\beta$ at the base are the base density and magnetic field intensity. Therefore by analyzing the beta profile we can interpret what is happening in our simulations for different input parameters. 

\begin{figure}
	\includegraphics[width=\columnwidth]{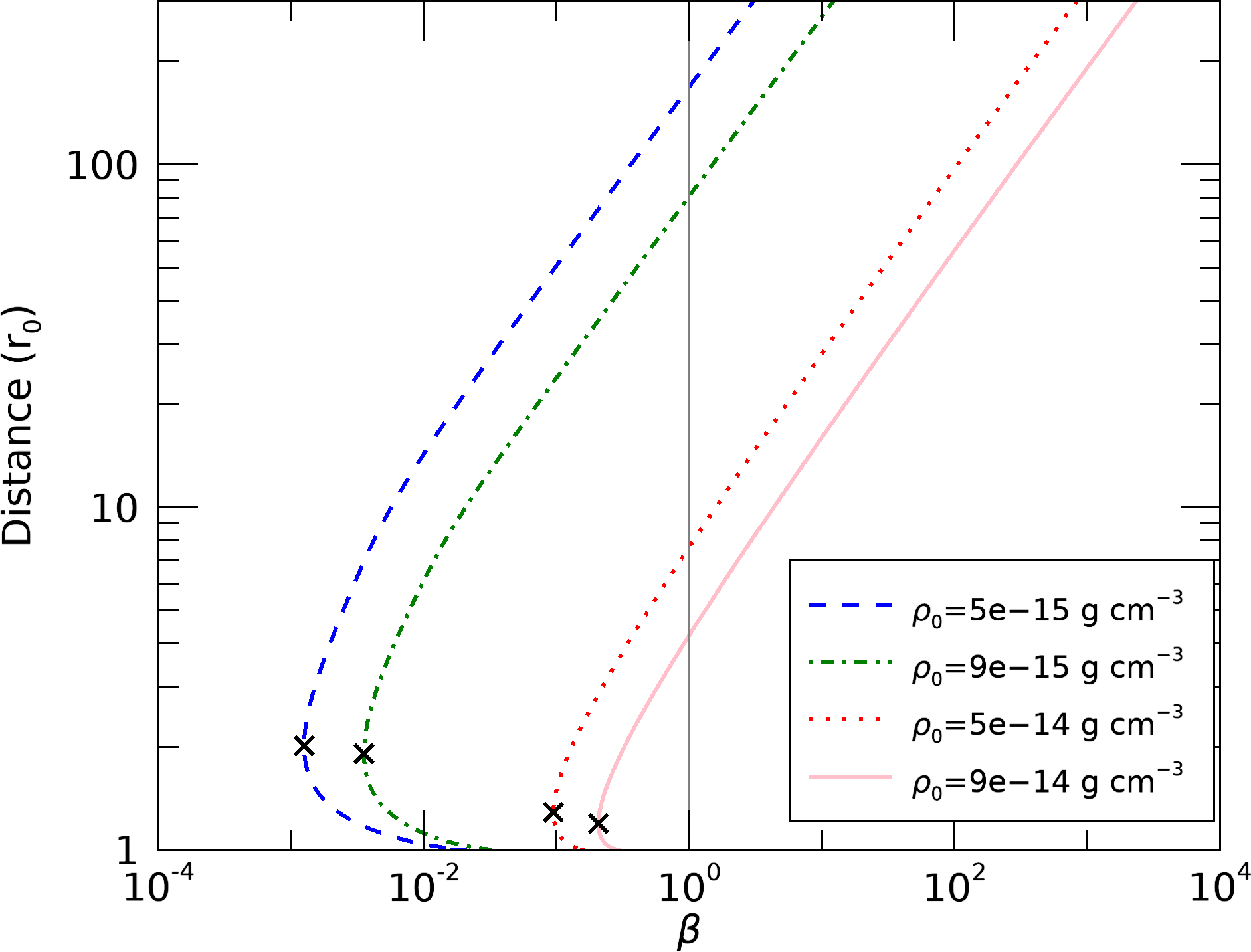}
    \caption{Plasma-$\beta$ as function of the distance for a constant magnetic field of 5\,G and different values of base density. The plot shows that the winds with lower densities (green-dash dotted and blue-dashed curves) are more magnetically dominated ($\beta<1$) and higher density winds (pink-solid and red-dotted curves) are more thermally dominated ($\beta>1$). The crosses denote the minimum of $\beta$, here defined as the base of the corona.}
    \label{fig:beta}
\end{figure}

\Cref{fig:beta} shows some selected $\beta$ profiles for $B_{0}=5$\,G and $\rho_{0} = 5$, 9, 50 and $90 \times 10^{-15}\,\text{g~cm}^{-3}$. In this plot, we see that smaller base densities (5 and $9 \times 10^{-15}\,\text{g~cm}^{-3}$) have $\beta < 1$ for nearly the whole wind. The wind only reaches $\beta>1$ for $r>100\,r_0$. For higher base densities (50 and $90 \times 10^{-15}\,\text{g~cm}^{-3}$), $\beta<1$ only for distances smaller than $10\,r_0$. These different profiles are due to a combination of different densities, temperature and magnetic field throughout the wind (see \Cref{fig:profiles}). These trends in the beta profiles demonstrate that low base density cases (low-$\beta$) are more magnetically dominated (magnetic field plays a major role in the wind) and high base density cases (high-$\beta$) are more thermally dominated (winds are thermally driven). This explains why temperature, velocity and density profiles for high-$\beta$ are not particularly affected by changes in magnetic field intensities. 

\subsection{Global trends of M dwarf winds}
\label{sec:wind-trends}
Here, we investigate the overall trends found in our simulations. To extract the global quantities of the wind, we use the fact that the values for temperature and velocity are nearly constant at large distance ($\gtrsim 50\,r_0$). We group simulations of same base density and, for each group, we extract local values of velocity, density and temperature at $r = 300\,r_0$, which represent the asymptotic terminal wind velocity ($u_\infty$), the density at large distances ($\rho_{\rm 300}$) and the ``isothermal'' (plateau) value of the temperature ($T_{\rm pl}$), respectively. Note that the density profile is not constant, but it continues to fall with $r^{2}$, following mass conservation of a constant-velocity wind. Within each $\rho_0$ group, there is a range of values of $u_{\infty}$, $\rho_{\rm 300}$ and $T_{\rm pl}$, due to different adopted $B_0$. To better identify the global trends, we average values of $u_{\infty}$, $\rho_{\rm 300}$ and $T_{\rm pl}$ for each group with same $\rho_0$.

\begin{figure}
    \includegraphics[width=\columnwidth]{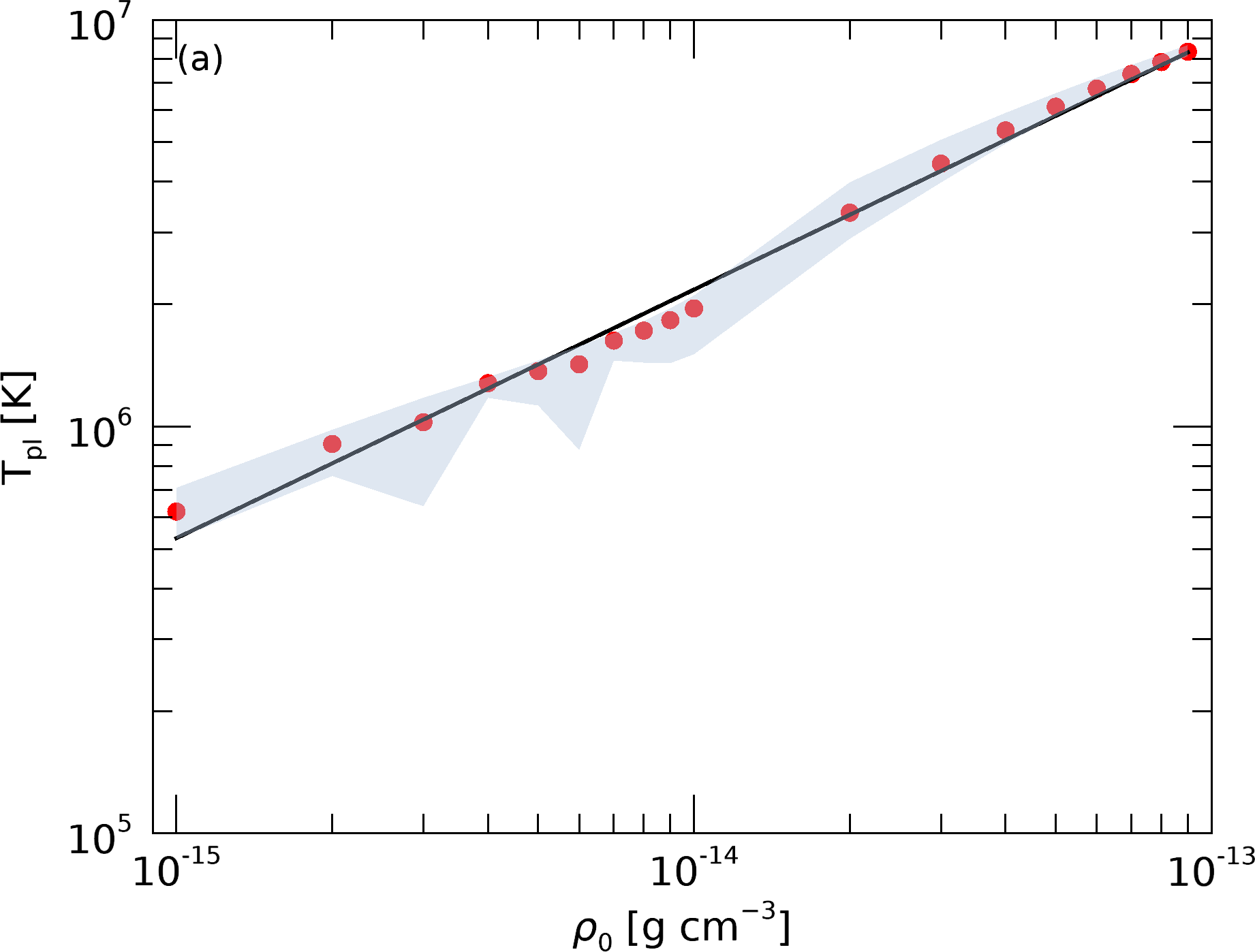}
    \includegraphics[width=\columnwidth]{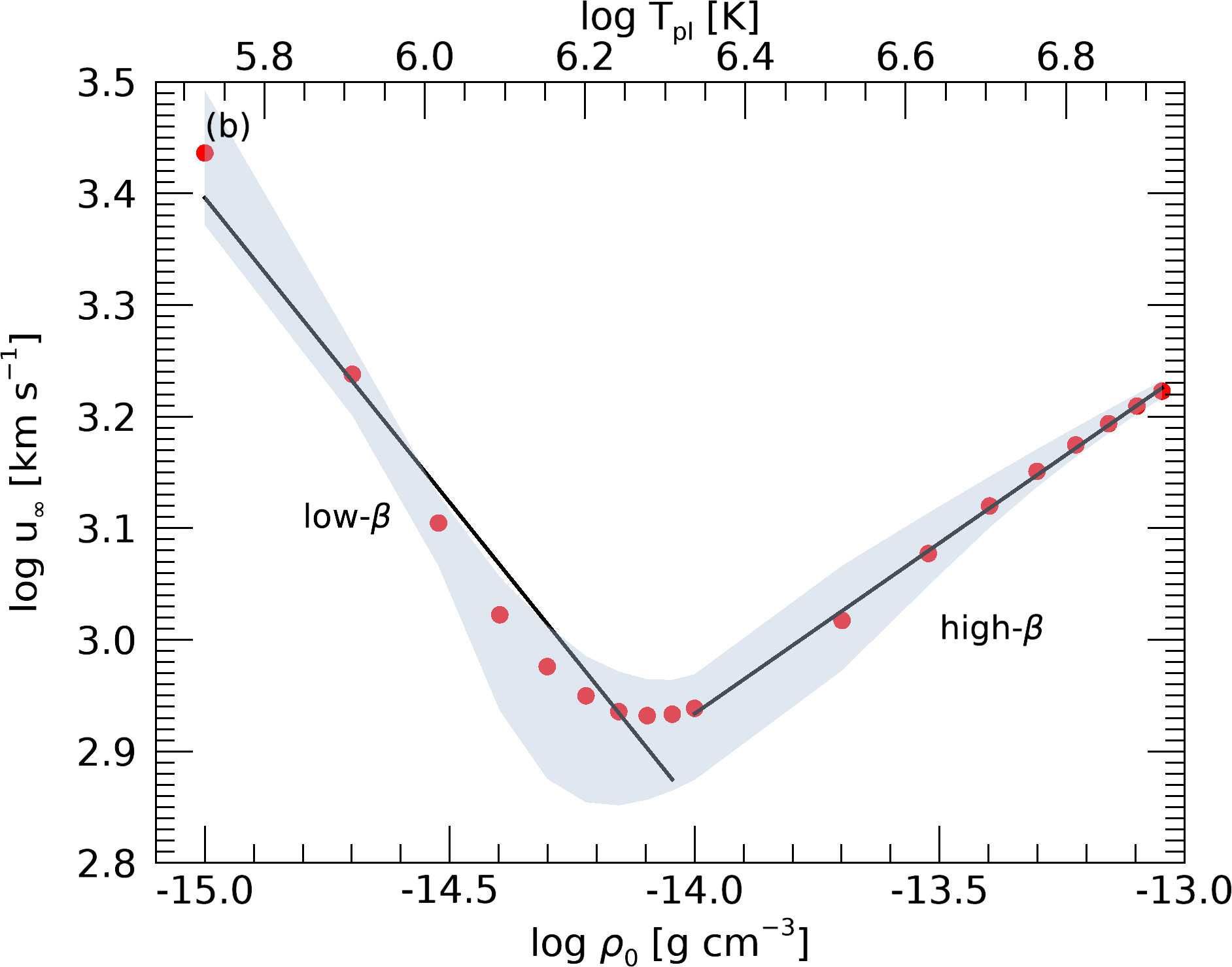}
    \includegraphics[width=\columnwidth]{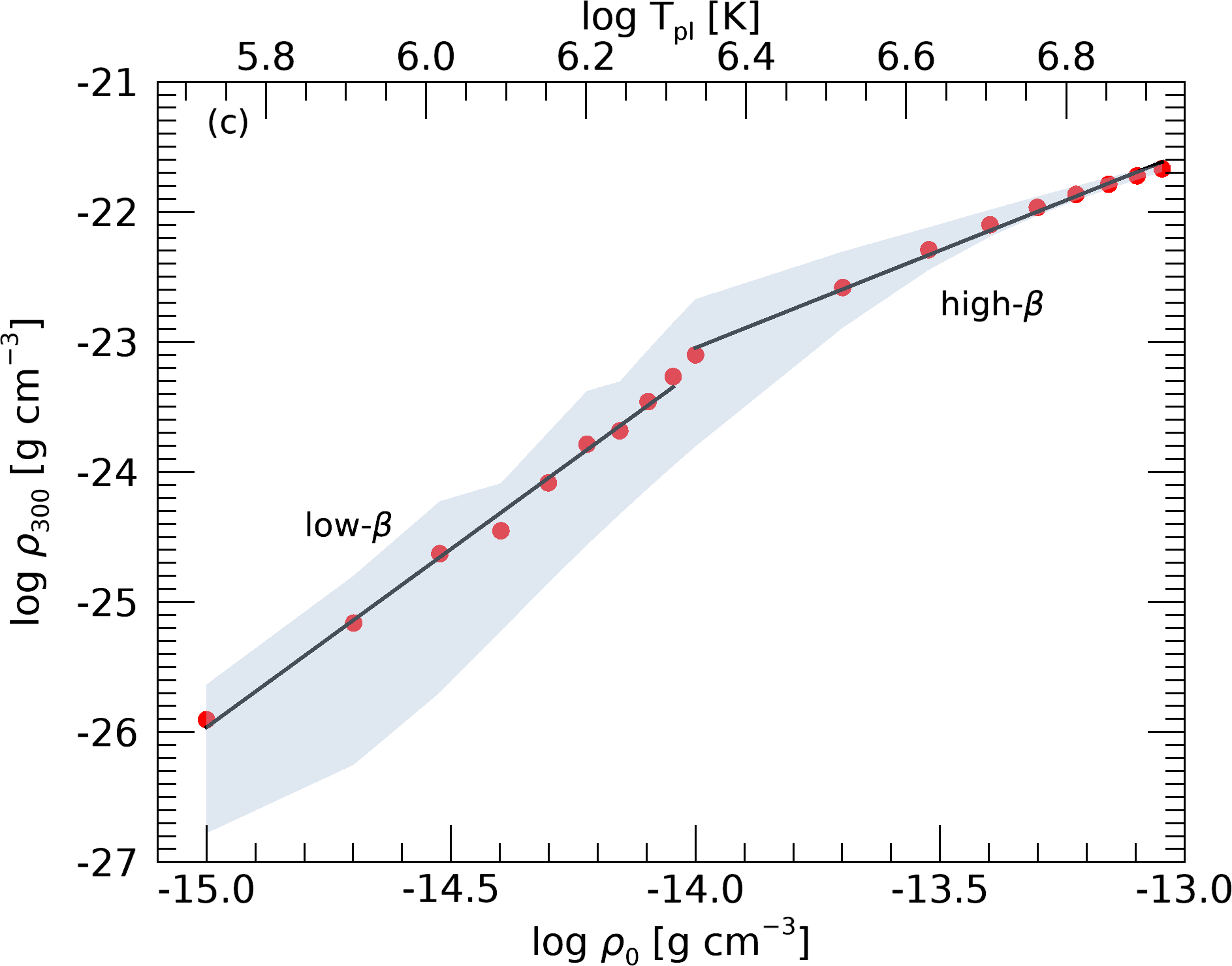}
  \caption{Given that for each $\rho_0$, there are multiple simulations with different $B_0$, we compute an average value (red points) of the (a) temperature plateau, (b) terminal velocity and (c) density at $r=300\,r_0$, for each group of simulations with the same $\rho_0$. We see two different regimes for terminal velocity and density at 300\,$r_0$, according to their values of plasma-$\beta$. The shaded areas in all plots represents the extreme values for each plotted quantity, that are due to different assumed $B_0$. The black lines are fits (\Cref{eq:temp,eq:velo,eq:den}).}
  \label{fig:general}
\end{figure}

\Cref{fig:general} shows the results we found for the general trends of the wind. The red points are the average values of $T_{\rm pl}$ and the solid line is the power-law fit. The shaded area in \Cref{fig:general}-a shows the range of the temperature plateau for different magnetic field values. The shaded area is larger for low-$\beta$ values, but overall we see that the averages (red points) are good representation of the different simulation parameters. The temperature plateau (\Cref{fig:general}-a) depends on the base density and can be described by a power law fit:
\begin{equation}
    T_{\rm pl} = (7.8 \pm 0.2) \times 10^{14}\,\rho_{0}^{0.61 \,\pm 0.01},
    \label{eq:temp}
\end{equation}
where $T_{\rm pl}$ is given in K and $\rho_0$ in g~cm$^{-3}$. The numbers in parentheses in \Cref{eq:temp} (and also in Equations \ref{eq:velo} and \ref{eq:den}) are the 1$\sigma$ uncertainties for each coefficient of the fit. In our models, winds with higher base densities use a higher fraction of the wave flux to heat the wind, thus we find that an increase in the base density also increases the temperature plateau.

The terminal velocity (\Cref{fig:general}-b) exhibits two different trends with base density. For low-$\beta$ range, the terminal velocity falls off with base density, while for high-$\beta$ range, the terminal velocity increases with base density. These two tendencies can be described by the power law fits:
\begin{equation}
    \begin{cases}
    u_{\infty} = (1.59 \pm 0.64)\times 10^{-5}\,\rho_{0}^{-0.55 \pm 0.04},\ \text{for low-}\beta \\
    u_{\infty} = (1.64\pm 0.07)\times 10^{7}\,\rho_{0}^{0.31 \pm 0.01},\ \text{for high-}\beta
    \end{cases}
    \label{eq:velo}
\end{equation}
where $u_{\infty}$ is given in km~s$^{-1}$ and $\rho_0$ in g~cm$^{-3}$. Here, we define low-$\beta$ for $\rho_0<10^{-14}\,\text{g~cm}^{-3}$ and high-$\beta$ for $\rho_0\geq 10^{-14}\,\text{g~cm}^{-3}$. The shaded area in \Cref{fig:general}-b shows the range of the terminal velocity for different magnetic field values. Opposite to what happens with the temperature plateau, the terminal velocity has a slightly large range with magnetic field, except for high-$\beta$ range. The terminal velocity shows a `V' shape profile with base density, where the average values of terminal velocities vary from approximately $850-2700\,\text{km~s}^{-1}$. In our simulations we can find the same terminal velocity for distinct values of wind temperature. Due to the fact that high-$\beta$ cases are thermally driven, the terminal velocity follows a simple Parker wind in where the higher the plateau temperature, the higher the terminal velocity. However, low-$\beta$ cases are more magnetically dominated and the wind speed becomes smaller with increase in base density for these cases. This occurs because the wind cannot be effectively accelerated due to the large quantity of material to lift up (larger inertia), which results in winds with lower terminal velocities. This is also seen in simulations by \citet{Suzuki2013}.

The density at $300\,r_0$ (\Cref{fig:general}-c) also has two different trends with base density, showing two different slopes for low and high-$\beta$ values.  These two trends can be described by the following power laws:
\begin{equation}
    \begin{cases}
    \rho_{300} = (1.44 \pm 1.17)\times 10^{15} \,\rho_{0}^{2.74 \pm 0.08},\ \text{for low-}\beta,  \\
    \rho_{300} = (8.35\pm 0.62)\times 10^{-3}\,\rho_{0}^{1.49 \pm 0.01},\ \text{for high-}\beta ,
    \end{cases}
    \label{eq:den}
\end{equation}
where $\rho_{300}$ and $\rho_0$ are given in g~cm$^{-3}$. Overall, the increase we see in $\rho_{300}$ is affected by the increase in temperature. Stellar wind density profiles can be approximated as an exponential decay closer to the star, with a certain scale height. Higher temperature winds have larger scale heights, thus a slow density decay with distance. This would explain why as we go to higher wind temperatures, the density at $300\,r_0$ remains larger (\Cref{fig:profiles}-c). The shaded area represents the range on the density at $300\,r_0$ for different magnetic field values. Similarly to the other plots, the shaded area is larger for low-$\beta$ and smaller for high-$\beta$ cases. The density at $300\,r_0$ is one of the most affected properties by the magnetic field variation, showing up to one order magnitude variation for the same $\rho_{0}$ in the low-$\beta$ regime.

Given the relation between $T_{\rm pl}$ and $\rho_0$ (\Cref{fig:profiles}-a), in Panels b and c of \Cref{fig:general}, we add a top axis indicating $T_{\rm pl}$ values.

\begin{figure}
	\includegraphics[width=\columnwidth]{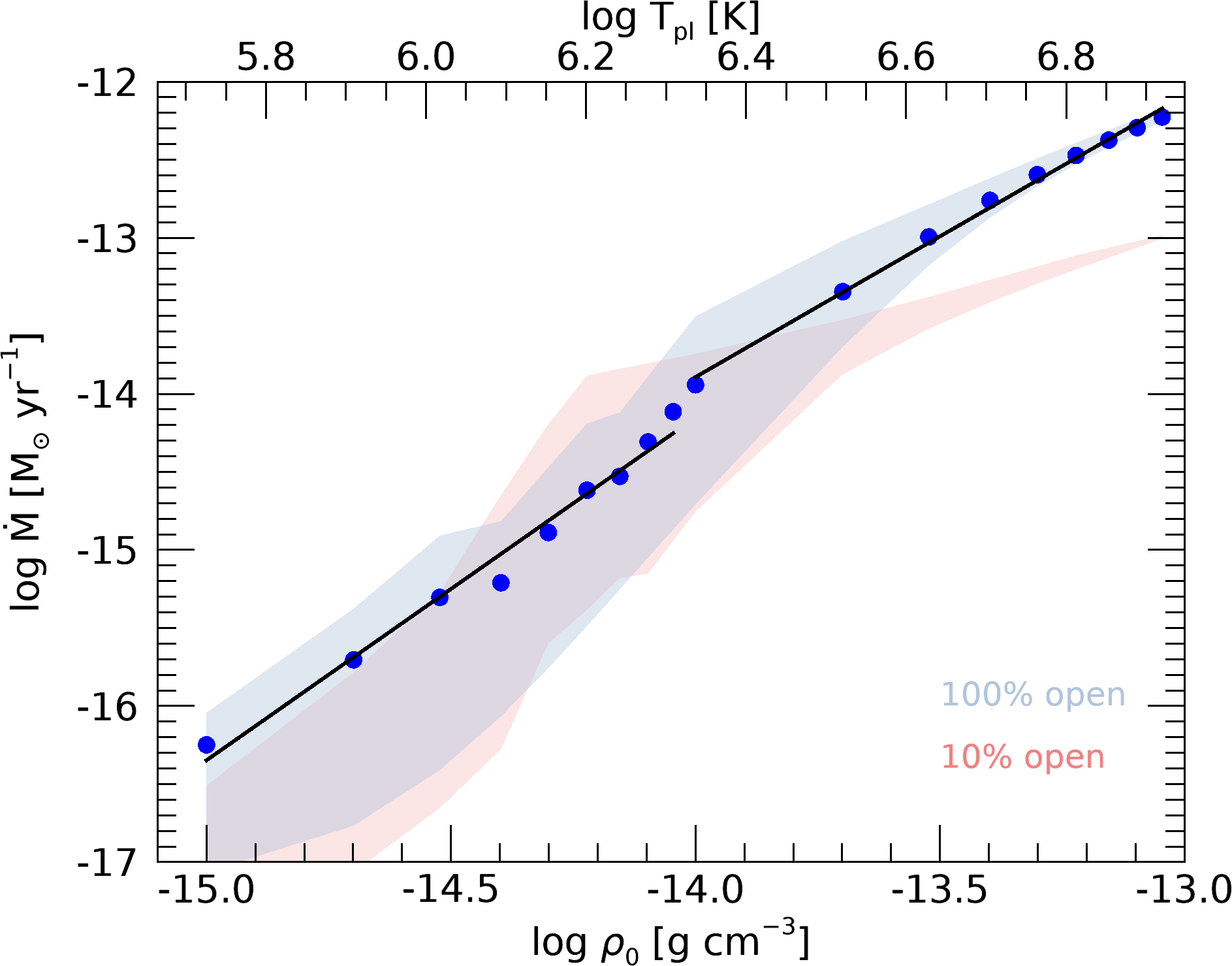}
    \caption{Mass-loss rate as function of base density and temperature plateau. The two straight lines are represented by \Cref{eq:mdot}. The shaded area represents the minimum and maximum values of mass-loss rate for each value of base density. The blue shaded area represents the model with 100\% open magnetic field lines and the red shaded area represents the model with only 10\% open magnetic field lines (see \Cref{sec:input}).}
    \label{fig:rho_mdot}
\end{figure}

The mass-loss rate can be calculated assuming spherical symmetry as
\begin{equation}
    \dot{M}=4\pi r^2 u\rho.
    \label{eq:massloss}
\end{equation}
Since the mass-loss rate depends of the velocity and the density, as a result it is possible to determine the trend of $\dot{M}$ with $\rho_{0}$. \Cref{fig:rho_mdot} presents the relation between mass-loss rate, base density and temperature plateau (grey shaded area). The solid lines are given by the equations:
\begin{equation}
    \dot{M}\propto \rho_{300}u_{\infty}\propto 
    \begin{cases}
    \rho_0^{2.19},\ {\text{for low-}}\beta\\
    \rho_0^{1.80},\ {\text{for high-}}\beta
    \end{cases}
    \label{eq:mdot}
\end{equation}
\Cref{eq:mdot} comes from the combination of \Cref{eq:velo,eq:den}. Even though the terminal velocity and the density at $300\,r_0$ have two different trends with $\rho_{0}$, the mass-loss rate increases with $\sim$ base density squared. Given the linear dependence of $\dot{M}$ with density (\Cref{eq:massloss}), why is $\dot{M}\propto \rho_{0}^{2}$? This is because the initial velocity of the wind, i.e., the one that is required for the wind to pass through the Alfv\'{e}n radius, has a linear relation with base density. I.e., the denser the wind, the larger is its required initial velocity.

\begin{figure}
	\includegraphics[width=\columnwidth]{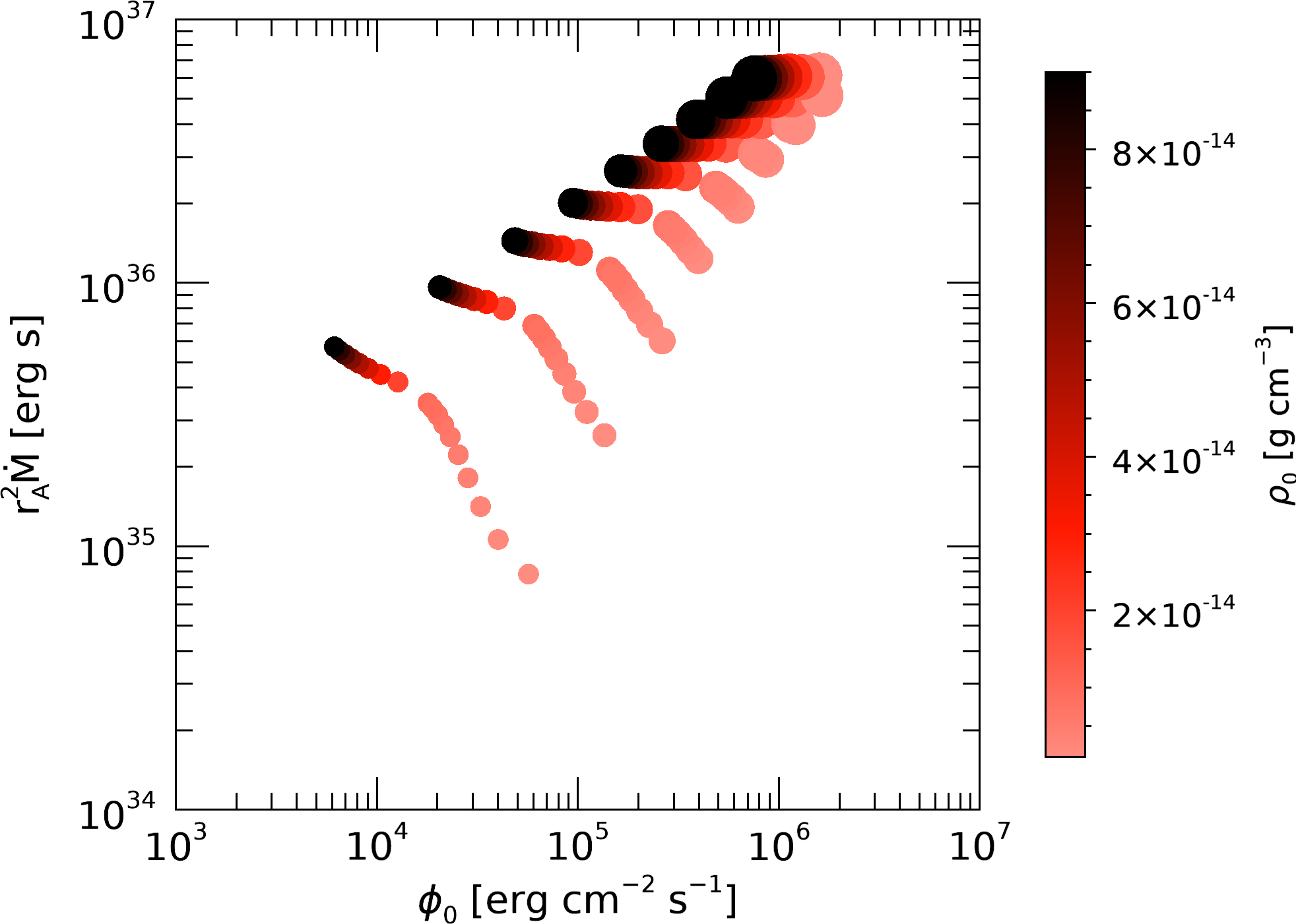}
    \caption{The angular momentum-loss rate, given by the proxy $r_A^2\dot{M}$, as a function of wave flux colour-coded according to base density. The symbol sizes are associated with magnetic field intensity, which varies from 1 (left set of points) to 10\,G (top right set of points). Angular momentum loss rates are larger for larger wave fluxes.}
    \label{fig:jdot}
\end{figure}

The wind achieves the Alfv\'{e}n velocity at the Alfv\'{e}n radius $r_A$. The Alfv\'{e}n radius determines, along with the mass-loss rate and rotation rate of the star, the amount of angular momentum that is carried away by the stellar wind. The angular momentum-loss rate is
\begin{equation}
    \dot{J} \propto \Omega_{\star}r_A^2\dot{M},
    \label{eq:jdot}
\end{equation}
where $\Omega_{\star}$ is the stellar rotation rate. The angular momentum is important to explain how the observed rotation periods of the stars change as they age \citep{Matt2015, Johnstone2015}. In our simulations, we do not consider rotation, so in \Cref{fig:jdot}, we show $r_A^2\dot{M}$, a proxy for the angular momentum-loss rate, as a function of base wave flux.

Rotation can alter the position of the Alfv\'{e}n radius and the mass-loss rate, but these parameters are more affected in the case of fast rotation. From \Cref{fig:jdot}, we see that for a given value of base density, the angular momentum-loss rate is higher for higher wave fluxes. The magnetic field intensity is represented by the size of the symbols. Given that $\phi_{A, 0} \propto \epsilon_{0} v_{A, 0} \propto (\delta B^{2})_{0} B_{0}/\sqrt{\rho_{0}}$, higher values of magnetic fields show higher values of wave flux, when the density is kept constant. For a given value of magnetic field, the angular momentum-loss rate decreases with wave fluxes, this trend is more evident for small $B_0$. In cases with higher $B_0$ the angular momentum-loss rate have a, roughly constant value, regardless of $\phi_{A, 0}$. 

\section{Applications of our model}
\label{sec:applic}
\subsection{The extension of the chromosphere in inactive/moderately active M dwarfs}
\label{sec:chrom}
The chromospheric size of the present Sun is less than 1\% of the solar radius \citep{Aschwanden2001,Suzuki2013}. In contrast, \citet{Czesla2012} showed, by using the Rossiter-McLaughlin effect, that a younger and active sun-like star, CoRoT-2A, could have a larger chromosphere extending to 16\% of the stellar radius. \citet{Suzuki2013} also analysed the time evolution of the height of the chromosphere for young solar-like stars and found that the size of the chromosphere is time dependent and varies from 10\%-20\% of the stellar radius. They define the top of the chromosphere as the distance where the temperature is $T=2\times 10^{4}\,\text{K}$. We use a different definition here, as we present below. 

We start our simulation in the chromospheric region and we assume by simplicity that the top of the chromosphere also defines the base of the corona. In the Sun, the base of the corona starts when the beta parameter (Eq.~\ref{eq:beta}) reaches a local minimum \citep{Gary2001, Aschwanden2001}. Below this local minimum, the photosphere has a high-plasma $\beta$, which reaches values of up to 100 (for a magnetic field of $\sim$ kG). Above this local minimum, the plasma-$\beta$ increases towards the corona. For the Sun, the minimum of plasma $\beta$ happens at $\beta \simeq 0.01$, at a height of $\sim 0.003\
R_{\odot}$ \citep{Aschwanden2001}.

We use the same idea here to define the top of the chromosphere/base of the corona in our simulations. In \Cref{fig:beta}, the crosses indicate the position of the local minimum (and therefore the base of the corona) for two cases with low base density (dashed and dash dotted curves) and two cases with high base density (dotted and solid curves). In \Cref{fig:beta}, for low-$\beta$ cases the local minimum of $\beta$ is around $2\,r_0$, while for the high-$\beta$, the local minimum of $\beta$ is $\sim 1.2\,r_0$. Overall, for all our simulations, we find that low-$\beta$ cases have the base of the corona in between 1.7 and 2.7\,$r_0$, while for high-$\beta$ cases, the base of the corona is in between 1.2 and 2.2\,$r_0$.

\begin{figure}
	\includegraphics[width=\columnwidth]{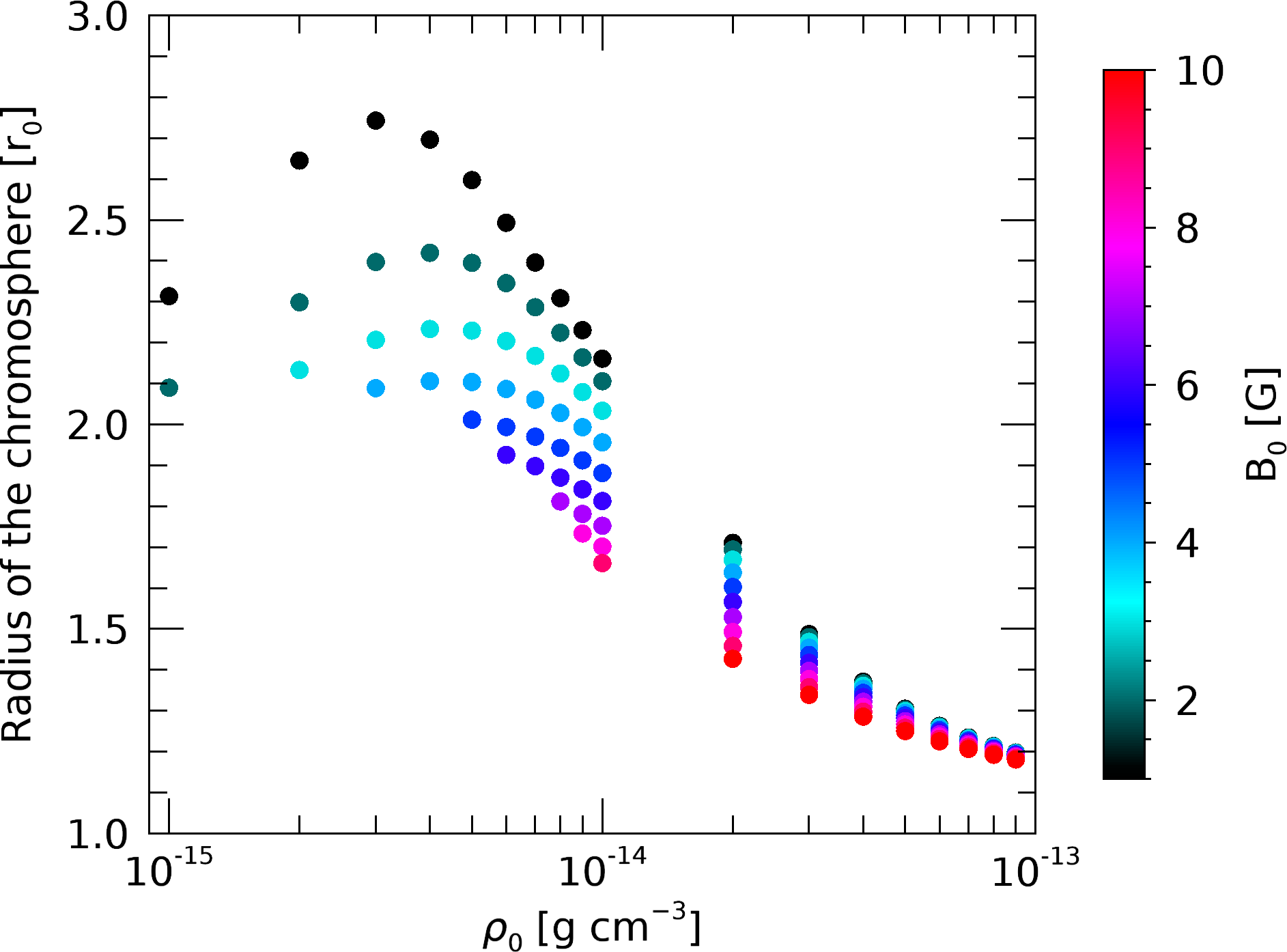}
    \caption{Extension of the chromosphere as function base density color-coded according to magnetic field intensity.}
    \label{fig:chrom}
\end{figure}

This process to define the base of the corona also gives us an estimate of the size of the chromosphere. \Cref{fig:chrom} shows the extension of the chromosphere as a function of base density where colour represents the magnetic field intensity. The larger the value of base density (high-$\beta$ range), the smaller is the chromosphere of the star. Winds with higher magnetic field intensities show smaller chromospheres. For low-$\beta$ range, we observe a large scatter in the chromospheric size.

From our simulations, we estimate that the size of the chromosphere is around 18\% -- 174\% of the stellar radius, depending on our inputs. The extension of the chromosphere has a very wide range of values in our simulations and is considerably larger than that observed by \citet{Czesla2012} and derived by \citet{Suzuki2013} in the context of solar-like stars. The difference between the results of \citet{Suzuki2013} and ours can be due to the different types of stars and/or definitions of the top of the chromosphere used by each work. In our cases, the temperature at the top of the chromosphere vary from 0.6 to 3.5\,MK, and, in theirs, it is assumed to be $2\times 10^{4}\,\text{K}$. They also have a transition region, which is not defined in our simulations.

\subsection{The wind of the planet-hosting star GJ 436}
\label{sec:star}
Cool dwarf stars, especially when more active, emit in X-rays. Here, we follow the work of \citet{Suzuki2013}, and use our wind models to estimate the X-ray emission of the planet-hosting star GJ 436. This star has measurements of both X-ray luminosity and mass-loss rate. GJ 436 is an M2.5 dwarf star, located at 10.14\,pc and host to an exoplanet GJ 436b at 0.0287\,au (about 14.1\,$r_0$). Based on XMM-Newton EPIC-pn spectrum of GJ 436, \citet{Ehrenreich2015} reported an X-ray flux of $4.6 \times 10^{-14}\,\text{erg~s}^{-1}\text{cm}^{-2}$ in the 0.12 -- 2.48\,keV band, resulting in an X-ray luminosity of $5.7\times 10^{26}\,\text{erg/s}$. 

By assuming that the radiative losses in the chromosphere is proportional to the X-ray luminosity, we can estimate the luminosity from the radiative losses ($P_r$): 
\begin{equation}
    L_X=\int P_r dV
\end{equation}
where $dV$ is the volume element. Here we perform the integral over $4\pi r^2 dr$, from $1<r<300\,r_0$, but note that only the inner region of the wind, within $2\,r_0$ contribute significantly to $L_X$. We note however that this underestimates the true X-ray luminosity of the star. Similar to the Sun, we expect that winds of M dwarfs are X-ray dark as they flow along open flux tubes (coronal holes) and are X-ray bright inside closed-field line regions. As in our simulations we only consider open flux tubes (i.e., the wind region), the observed X-ray luminosity is used as an upper limit to rule out certain simulations in our parameter space. With this, we can place an upper limit to the mass-loss rate of GJ 436. This is presented next.

\begin{figure}
    \includegraphics[width=\columnwidth]{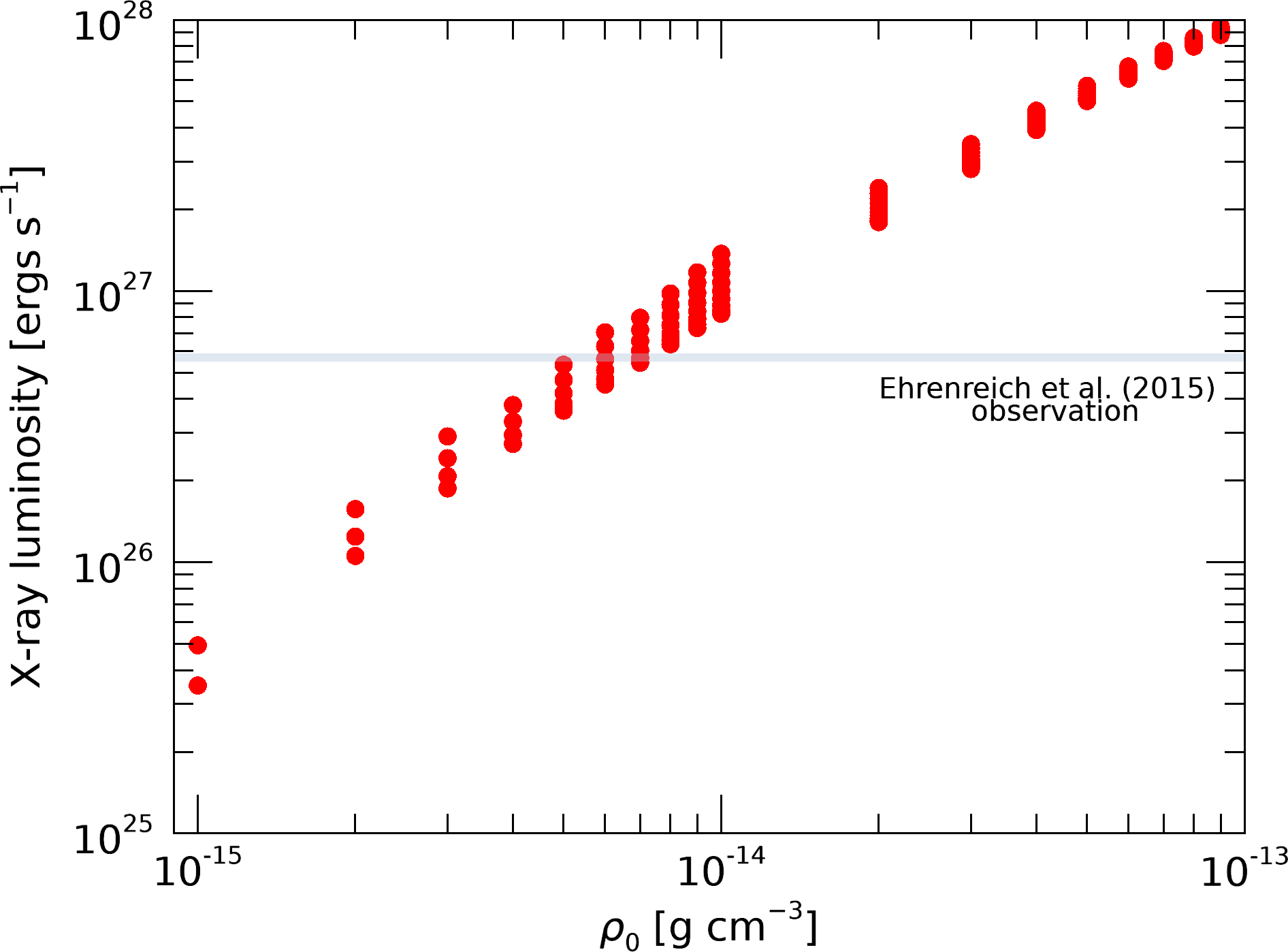}
    \includegraphics[width=\columnwidth]{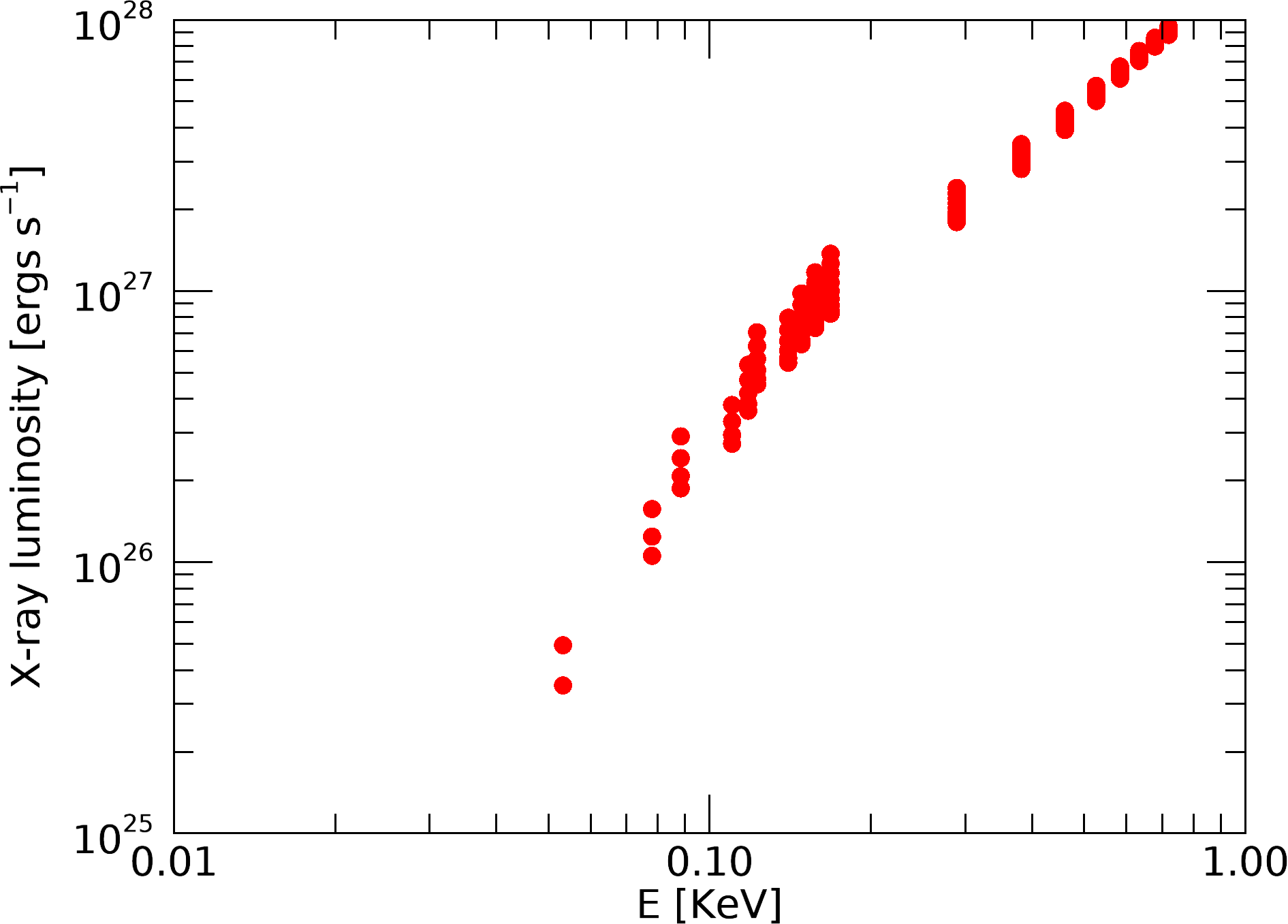}
  \caption{X-ray luminosity from radiative losses as function of the base density (top) and energy (bottom). The X-ray luminosity observed for GJ 436 is represent by the shaded area with the respective base density.}
  \label{fig:lumi}
\end{figure}

\Cref{fig:lumi} shows our computed X-ray luminosity as a function of the base density (top) or energy $E=k_{B}T_{\rm pl}$ (bottom) for all our simulations. We get an X-ray luminosity ranging from $3.5\times 10^{25}$ -- $9.5\times 10^{27}\,\text{erg~s}^{-1}$ and energy range from 0.05 -- 0.7\,keV (corresponding to a wavelength ranging from 17 to 234\,\AA). The energy bound includes the radiation in the ultraviolet range and also in the X-ray range.

The observed luminosity is marked in \Cref{fig:lumi} by a horizontal line. We see that the models with base densities smaller than $7\times 10^{-15}\,\text{g~cm}^{-3}$ would give rise to luminosities similar or smaller to the observed one. These models produce mass-loss rates smaller than $7.6\times 10^{-15}\,M_{\odot}~\text{yr}^{-1}$, with Alfv\'{e}n radius varying from 23$\,r_0$ -- 75$\,r_0$. This puts the planet orbiting at a distance below the Alfv\'{e}n radius, in a sub-Alfv\'{e}nic region. Because of the sub-Alfv\'{e}nic interaction, energy can be transported back to the star, potentially causing star-planet interaction signatures on the star \citep{Saur2013}.

\citet{Vidotto2017} used modelling of stellar wind interactions with upper planetary atmosphere of the warm-neptune GJ 436b to derive the global characteristics of the wind of GJ 436. Using this approach they estimated the mass-loss rate to be (0.5 -- 2.5)$\times 10^{-15}\,M_{\odot}~\text{yr}^{-1}$, which is within our predictions. However, inspite of the mass-loss rate agreement, our models predict a local velocity that do not match with their values. While \citet{Bourrier2016} reported a local velocity of 69 -- 91\,km/s at the orbit of the planet, our models give higher velocities with values $<$ 800\,km/s. Our local densities are < $2.8\times 10^{-21}$\,g~cm$^{-3}$, in agreement with theirs (1.34 -- 7.02) $\times 10^{-21}$\,g~cm$^{-3}$. Overall, the local wind velocity has a higher value in our work, but density has a similar value, and the combined values give a similar mass-loss rate to \citet{Vidotto2017}. By modelling the wind of GJ 436 as an isothermal wind, \citet{Vidotto2017} found that the Parker wind cannot satisfy simultaneously the wind density, temperature and velocity reported in \citet{Bourrier2016}. They suggested that this could be due to a different, or additional, acceleration mechanism for the wind (i.e., other than the thermal forces), such as, for example, the Alfv\'{e}n-wave pressure force. Similar to their findings, our model cannot reproduce simultaneously the wind density, velocity and temperature reported in \citet{Bourrier2016}. 

By modelling the Lyman-$\alpha$ transits, other works have also derived the properties of the wind of GJ 436, as it interacts with the warm Neptune GJ 436b. All these results are summarised in \Cref{tab:output}. It is surprising to see overall agreement between all these different models. All the works have densities of the same order of magnitude. Local wind velocities are also all in the same ballpark, although, except for Villarreal D'Angelo et al. (in prep), other models predict  a factor of 4 to 8 smaller velocities than ours. Mass-loss rates of all these works are also of similar magnitude, except for \citet{Kislyakova2019}, who found $\dot{M}$ higher than the other works. Maybe the largest disagreement is on the temperature values: our values are  higher by a factor of a few than the values found by other works. 

\begin{table*}
 \caption{Comparison of local stellar wind properties of GJ 436 at the position of GJ 436b for different works.}
 \label{tab:output}
 \begin{tabular}{ccccc}
  \hline
  Velocity (km/s) & Density ($10^{-21}$ g~cm$^{-3}$) & Temperature (MK) & $\dot{M}$ ($10^{-15} M_{\odot}~\text{yr}^{-1}$) & 
  \\
  \hline
  < 800 & < 2.8 & < 1.7 & < 7.6 & this work\\ 
  \\
  69 -- 91 & 1.34 -- 7.02  & 0.36 -- 0.46 & 0.5 -- 2.5 & \citet{Vidotto2017}\\
  110 & 3.4 & 0.41 & 35 & \citet{Kislyakova2019}\\
  170 & 6.7 & 0.6 & 4 & set No 8 of \citet{Khodachenko2019}\\
  470 & $0.5$ & 0.17 & 2 & Villarreal D'Angelo et al. (in prep)\\
  \hline
 \end{tabular}
\end{table*}

Still, the overall agreement indicates that planetary transits can be used as a way to study stellar wind properties, as proposed by \citet{Vidotto2017}. We note however, that, different models, like the Alfv\'{e}n-wave driven wind and Parker wind models, can show similar properties at the orbital distances of exoplanets (\Cref{sec:parker}). As a consequence, by using only planetary transit observations, it is difficult to distinguish between different models. Thus, it is more likely that models would only be able to derive some global characteristics of the wind (like mass-loss rate), but not the detailed physics of wind acceleration.

\section{Discussion}
\label{sec:discu}
\subsection{Comparison between the Alfv\'{e}n-wave driven wind (AWDW) and the Parker wind (PW) models}
\label{sec:parker}
It is interesting to investigate how our results compare to the most commonly-adopted stellar wind model, namely the isothermal, PW model. This has been done, for example, for solar wind simulation \citep{2017ApJ...835..220C}, although here we use a different comparison method. In our comparison, we calculate the isothermal wind solution for each of our simulations. One free parameter in the PW is the temperature. Given that our simulations reach a temperature plateau, we use this temperature plateau as an input for our PW simulations. While the base density plays an important role in the temperature profile in the AWDW simulations, this is not the case for a PW. Since the isothermal wind equations are independent of the density, the base density itself in a PW is a scaling factor for the mass-loss rate. Below we present a scheme of how we use the outputs of AWDW as a input for the PW. 
\begin{equation*}
    \rho_{0\, {\rm (AWDW)}} \Rightarrow 
    \begin{cases}
    \langle T_{\rm pl}\rangle\\
    \langle\rho_{\rm cor}\rangle
    \end{cases}
    \Rightarrow \text{Parker Wind} \Rightarrow 
    \begin{cases}
    u_{\infty \, \rm (PW)}\\
    \dot{M}_{\rm (PW)}
    \end{cases}.
\end{equation*}
For a given set of simulations with same $\rho_0$ in our AWDW model, we extract values of average temperatures plateau and coronal base densities. These are then used as input for a PW simulation, which results in values of terminal velocity and mass-loss rate.

Unlike the AWDW simulation, the PW simulation starts at the corona, where the temperatures have already reached around a million K, therefore the base density needs to be chosen accordingly.  We inspect the results of the AWDW to get the density of the corona for each simulation (see \Cref{sec:chrom} for definition of the corona). This density at the base of the corona was used as the input density for the PW.

\begin{figure}
	\includegraphics[width=\columnwidth]{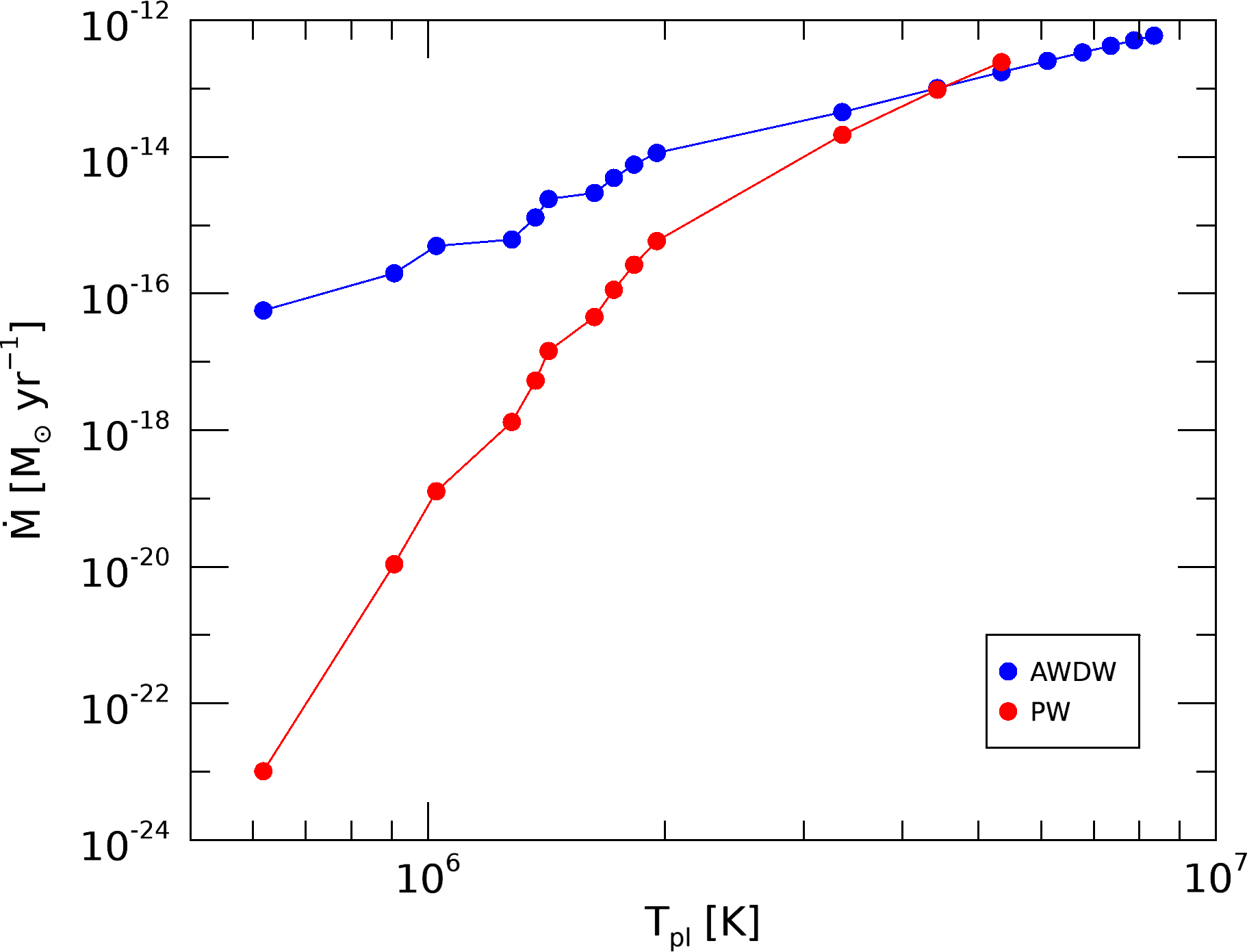}
    \caption{Mass-loss rate using Alfv\'{e}n-wave driven wind (blue dots) and Parker wind (red dots). The PW mass-loss rate is underestimated for winds with low temperature plateau, and it is in good agreement with AWDW for high temperatures. The PW do not extend out to the same temperature range as the AWDW because the solution does not pass trough the sonic point for those $T_{\rm pl}$ values.}
    \label{fig:mdot}
\end{figure}

\Cref{fig:mdot} shows the mass-loss rates, as calculated by \Cref{eq:massloss}, for the PW (red dots) and AWDW (blue dots). At the $T_{\rm pl}\gtrsim 10^{6.34}$\,K, both wind mechanisms give similar values for mass-loss rate. This is because the velocity and density at large distances are similar in both models (see \cref{sec:appendix}). In contrast, at the $T_{\rm pl}< 10^{6.34}$\,K, the PW underestimates mass-loss rate by several orders of magnitude. The difference is particularly high ($\geq 10^{4}$ times) for cases with base density values smaller than $4\times 10^{-15}\,\text{g~cm}^{-3}$.

In conclusion, the PW is a good representation of the AWDW for high-$\beta$, where the wind is thermally dominated. However, the PW can underestimate the terminal velocity and density, and thus mass-loss rate, for the low-$\beta$ cases, where the wind is magnetically dominated.

\subsection{Different parameters of the model}
\label{sec:input}
In the AWDW model, we have a considerable number of free parameters: $\rho_0$, $B_0$, $T_0$, $L_0$, $\sqrt{\langle\delta B_0^2\rangle}$, damping type and two parameters to describe the flux tube geometry and coverage; in opposition to PW model which needs only $T_0$ and $\rho_{cor}$. We investigate now how these parameters affect the structure of our winds.

\subsubsection{Flux tube geometry}
\label{sec:geom}
\begin{figure}
\includegraphics[width=\columnwidth]{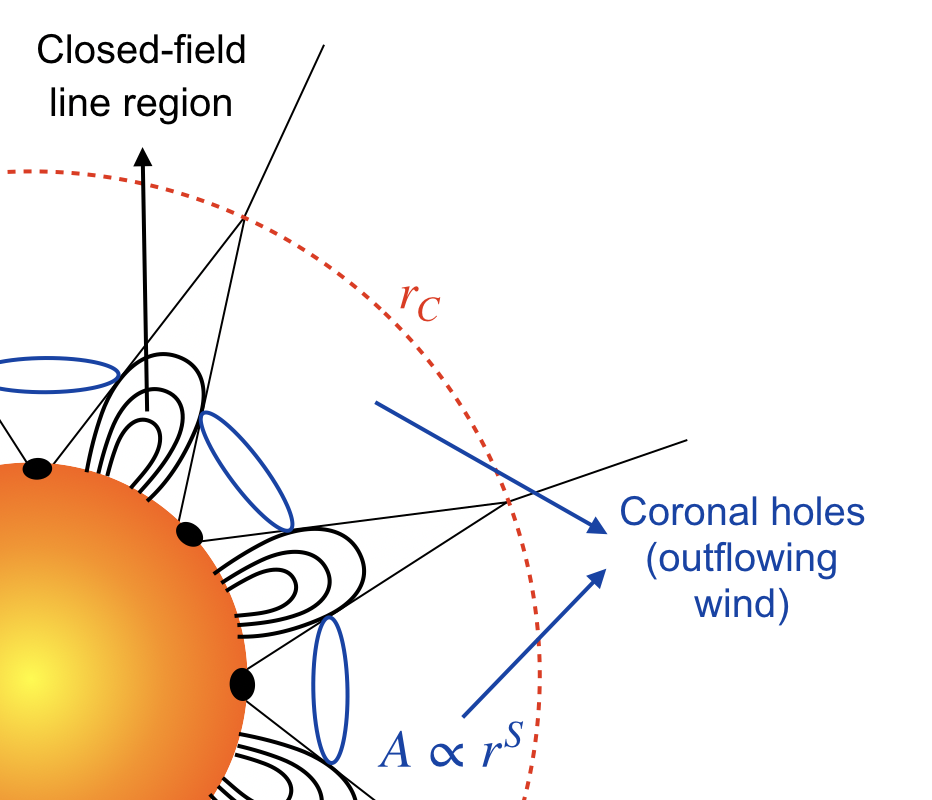}
 \caption{Sketch of a stellar atmosphere showing regions with closed-field lines and open-field lines (coronal holes). The coronal holes have super-radial flux tubes (area $A\propto r^S$, where $S$ in the super radial expansion factor) until the point $r_C$, beyond which the flux tubes become radial.}
 \label{fig:sketch}
\end{figure}

In order to investigate how the flux tube geometry and coverage affect our results, we  run a set of 125 simulations with only 10\% of open magnetic field line configuration. To implement this configuration we use a filing factor of open flux tubes, defined as $f_0=\frac{1}{F_0}=0.1$, which defines the open flux tubes coverage at the stellar surface. The open flux tubes have a super-radial expansion until a distance $r_c$, which defines the extension of the closed-field lines \citep{Vidotto2006}. See sketch in \Cref{fig:sketch}. 

For a given $S>2$, $r_c$ \citep{Kuin1982, Vidotto2006} is defined by
\begin{equation}
    F_0= \frac{\Omega_c}{\Omega_0}=\frac{A(r_c)/r_c^2}{A(r_0)/r_0^2}=\left(\frac{r_c}{r_0}\right)^{S-2},
\end{equation}
where $\Omega_c$ and $\Omega_0$ are the solid angle at $r=r_c$ and $r=r_0$, respectively, $S$ the super radial expansion exponent and $A$ is the area. We chose $S=4.095$ because it gives rise to an extension of the closed field line region of $r_c=3\,r_0$, which is similar to the value observed in the solar wind (i.e., above $r_c$ the magnetic field is purely radial). The area of the flux tube is defined as
\begin{equation}
    A(r)= 
    \begin{cases}
    A(r_0)(r/r_0)^S,\ \text{if} \ r\leq r_c\\
    A(r_0)(r_c/r_0)^S (r/r_c)^2,\ \text{if} \ r> r_c.
    \end{cases}
\end{equation}

The new configuration (with 10\% open magnetic field lines) shows a similar value of temperature plateau when compared with simulations with 100\% open magnetic field lines. The density at $300\,r_0$ is similar for low-$\beta$ cases but smaller for high-$\beta$ cases, around one order of magnitude. The terminal velocity is similar for high-$\beta$ cases, but for low-$\beta$, $u_{\infty}$ is considerably smaller, around half of the value. This can be understood by using an analogy of a pipe: if the aperture of the pipe is reduced, the velocity of the flow through the pipe increases. Similarly, if the aperture of the pipe is increased, the velocity of the flow goes down (assuming the same flux in both cases). The mass-loss rate have the same main trends as the $\rho_{300}$, as can be seen in \Cref{fig:rho_mdot} (red shaded area). In spite of changes in density and velocity, overall the mass-loss rates in both set of simulations are comparable.

\subsubsection{Thermal properties}
\label{sec:temp}
Back to our original geometry, we also run a set of simulations with a higher base temperature of $5\times 10^{4}\,\text{K}$ (2.5 times higher than the previous value). Overall, the temperature of the wind increases, but we found that a higher $T_0$ does not significantly affect the wind velocity profile. However, the mass-loss increases by around one order of magnitude when compared with a lower base temperature. This is the same problem as seen in PW models, namely, that the mass-loss rate is sensitive to the temperature of the wind.

\subsubsection{Properties of the waves}
\label{sec:wave}
As discussed before in \Cref{sec:star}, our velocity values at the orbit of the planet GJ 436b is much higher than the values reported by \citet{Bourrier2016}. In order to check if we could reproduce the low wind speed at GJ 436b orbit, we run our simulations with densities (6 -- 7)$\times10^{-15}\,\text{g~cm}^{-3}$ (which reproduces the luminosity value observed for GJ 436) but changing some properties of the waves, namely $L_0$ and $\sqrt{\langle\delta B_0^2\rangle}$. 

When we run the simulation with $L_0=0.01\,r_0$ (one order of magnitude smaller than our main simulations), the temperature profile decreases (the factor depends on the $B_0$ intensity). The velocity profile also shows a small reduction. For example, if we compare the profiles for $\rho_0=6\times10^{-15}\,\text{g~cm}^{-3}$ and $B_0=1$\,G, the velocity at the planet orbit is $u_p=430$\,km/s using $L_0=0.1\,r_0$ and $u_p=402$\,km/s using $L_0=0.01\,r_0$. As we can see, there is a decrease in the velocity, but it is still much higher than the value observed by \citet{Bourrier2016}.

When we run the simulation with $\sqrt{\langle\delta B_0^2\rangle}=0.01\,B_0$ (one order of magnitude smaller than our main simulations), the density decreases less than one order of magnitude, the temperature also decreases, but to a lesser extent. The velocity, goes down very significantly, with the decrease being stronger for higher $B_0$. For instance, for $\rho_0=6\times10^{-15}\,\text{g~cm}^{-3}$ and $B_0=3$\,G, the velocity at GJ 436b orbital distance is $u_p=711$\,km/s using $\sqrt{\langle\delta B_0^2\rangle}=0.1\,B_0$ and $u_p=348$\,km/s using $\sqrt{\langle\delta B_0^2\rangle}=0.01\,B_0$, which is about half of the previous value. The side effect of using a smaller value of $\sqrt{\langle\delta B_0^2\rangle}$ is that it affects directly the mass-loss rate resulting in a new value smaller by around one order of magnitude.

\medskip
From the discussion presented in \Crefrange{sec:geom}{sec:wave}, we conclude that, as a consequence of several free parameters, it might be possible to find a set of inputs that would reproduce the observations by \citet{Bourrier2016}. However, this is beyond the scope of this paper and, in this work, we focus on discussing general trends of our model.

\section{Conclusions}
\label{sec:conc}
In this paper, we investigated the general trends of winds of M dwarfs. Our goal was to derive the main properties of the winds, including properties that could be observationally tested (size of the chromosphere and lower limits to X-ray luminosities). For that, we investigated how stellar winds from M dwarfs are affected by variations in the magnetic field and density at the chromosphere. Overall, we performed more than 300 MHD simulations with Alfv\'{e}n wave energy fluxes spanning 4 orders of magnitude.

We classified our simulations in low-$\beta$ ($\beta$<1) and high-$\beta$ ($\beta$>1) regimes, which is related to the adopted values of base density (low and high, respectively). When the base density is larger, the temperature and density profiles are larger. The velocity profile has two different regimes: it decreases for low-$\beta$ and increases for high-$\beta$ with base density. To a lesser extent, the temperature, velocity and density profiles increase with magnetic field intensity (\Cref{fig:profiles}).

We calculated the mass-loss rate using \Cref{eq:massloss} and found that our mass-loss rates are proportional to $\rho_{0}^{2}$. The square dependency with base density is associated to the fact that input velocity, required for the wind to pass trough the Alfv\'{e}n radius, is higher for cases with higher base density. We also calculate $r_A^2\dot{M}$, which is proportional to the angular-momentum loss rate $\dot{J}$ (\Cref{eq:jdot}). We found that $\dot{J}$ increases overall with wave base flux (\Cref{fig:jdot}).

When compared to the Parker wind (PW), we showed that Alfv\'{e}n-wave driven wind (AWDW) model accelerates more quickly but both wind mechanisms reach a similar terminal velocity for high-$\beta$. The PW can underestimate density by several orders of magnitude when compared with the AWDW -- this feature is more accentuated for the low-$\beta$ regime and is a consequence of both the large-distance density ($\rho_{300}$) and terminal velocity ($u_{\infty}$) being underestimated in the PW model (\Cref{fig:mdot}). On the contrary, for high-$\beta$, both wind mechanisms give a similar mass-loss rate. This is due to the fact that the high-$\beta$ regime is thermally dominated. We conclude that, the PW is a good representation of the AWDW for high-$\beta$, where the wind is thermally dominated. However, the PW can underestimate the terminal velocity and density, and thus mass-loss rate, for the low-$\beta$ cases, where the wind is magnetically dominated.

As applications of our model, we use the local minimum of the plasma beta parameter to define the transition between the chromosphere and corona. We found that the size of the chromosphere for M dwarf stars is more extended than that of our present-day Sun. We found that M dwarfs can have a very wide chromosphere extending to 18\% -- 174\% of the stellar radius and is larger for the low-$\beta$ regime.

Assuming that the X-ray luminosity is proportional to the radiative losses in the chromosphere, we estimated the X-ray luminosity from our stellar wind models. We compared our results with the observed X-ray luminosity of GJ 436 to constrain its mass-loss rate to be $\dot{M}<7.6\times 10^{-15}\,M_{\odot}~\text{yr}^{-1}$, with local velocities smaller than 800\,km/s, local densities smaller than $2.8\times 10^{-21}$\,g~cm$^{-3}$ and local temperatures 1.4 -- 1.7\, MK. Overall, our results are in good agreement with works that use Lyman-$\alpha$ transits to constrain the properties of the stellar wind (\citealt{Vidotto2017, Kislyakova2019, Khodachenko2019}, Villarreal D'Angelo et al., in prep). This indicates that transmission spectroscopy of planetary transits coupled with models can be used as a way to study stellar wind properties \citep{Vidotto2017}.

\section*{Acknowledgements}
The authors acknowledge funding from the Provost's PhD Project Awards, without which this work would not have been possible. 
AAV acknowledges funding from the Irish Research Council Consolidator Laureate Award 2018. This project has received funding from the European Research Council (ERC) under the European Union's Horizon 2020 research and innovation programme (grant agreement No 817540, ASTROFLOW).


\bibliographystyle{mnras}
\bibliography{bib/reference.bib}



\appendix
\section{Further comparison between our wind models and a Parker wind}
\label{sec:appendix}
\begin{figure}
	\includegraphics[width=\columnwidth]{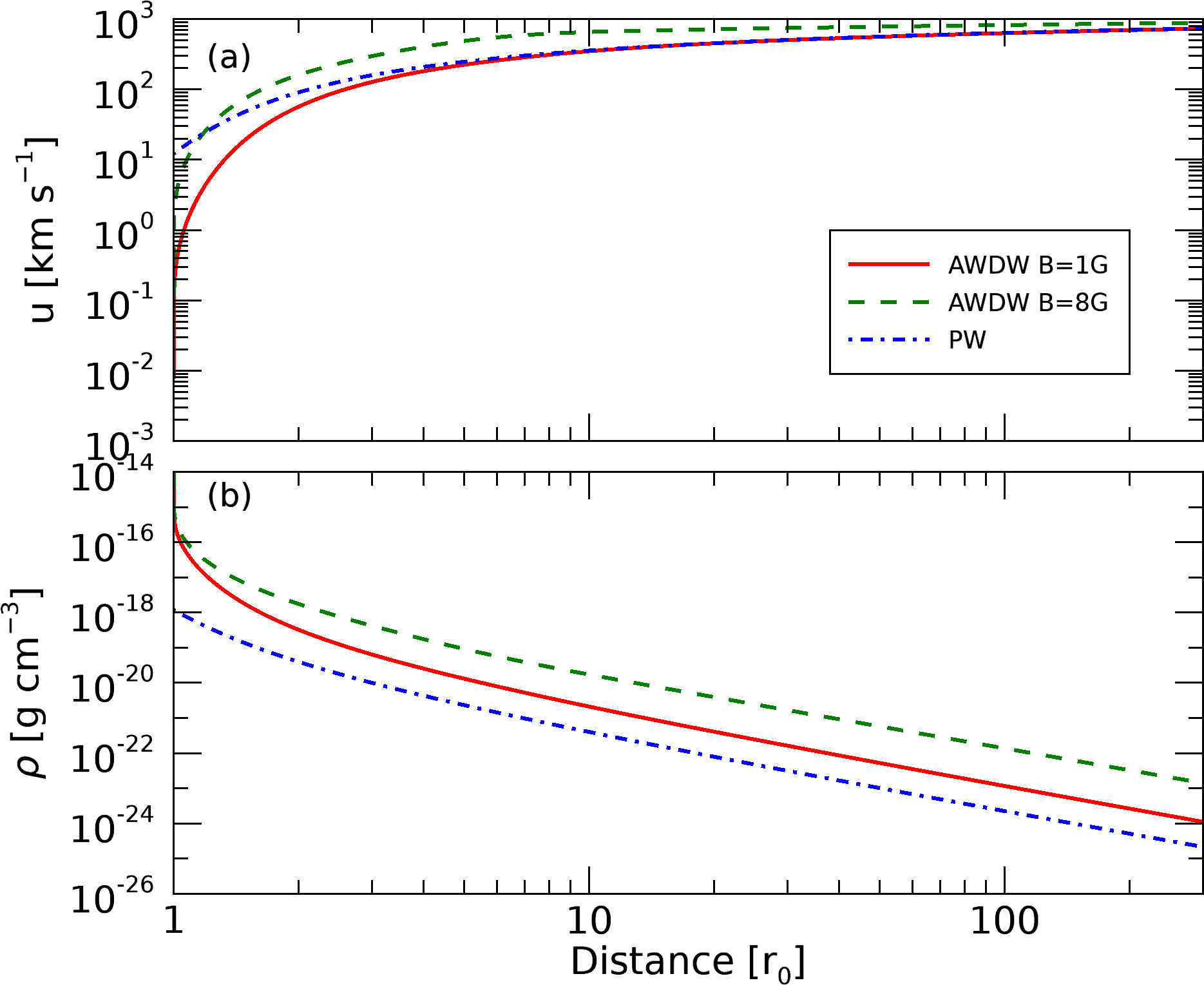}
    \caption{Comparison between the profiles for Parker wind (blue-dash doted curve) and Alfv\'{e}n-wave driven wind with B=1\,G (red-solid curve) and with B=8\,G (green-dashed curve). (a) velocity profile and (b) density profile.}
    \label{fig:comparison}
\end{figure}

Here we present further comparison between the AWDW and the PW, following \Cref{sec:parker}. We show in details the differences in the density and velocity profiles and the ratio of $u_{\infty}$ and $\rho_{300}$.
\Cref{fig:comparison} shows the comparison between the AWDW simulations and the PW simulations for a few selected cases: $\rho_0=9\times 10^{-15}\,\text{g~cm}^{-3}$ and $B_{0}$ = 1\,G (red-solid line) and 8\,G (green-dashed line). The main difference is that the AWDW accelerates more quickly than the PW, but both methods reach a similar terminal velocity. This is due to the fact that once the AWDW reaches $T_{\rm pl}$, and thus most of the wave energy has been deposited in the wind, the wind becomes thermally driven, similar to a PW. \Cref{fig:comparison}-a shows the velocity profile for both methods. It is interesting to note the similar decay of the three curves at large distance: this is the $r^{-2}$ decay of the density that is seen in all models. The density profile (\Cref{fig:comparison}-b) for the PW displays a more rarefied wind. Here, the coronal density of the PW was defined as the average density for all the cases with $\rho_0=9\times 10^{-15}\,\text{g~cm}^{-3}$. We could, in principle, have scaled the dashed-dot blue curve of the PW to match either the 1-G or the 8-G model, given that the density in the PW is a scaling factor. This would force the density at large distances to be the same in the AWDW and PW models, but then they would deviate from each other at small distances. 
   
\begin{figure}
    \includegraphics[width=\columnwidth]{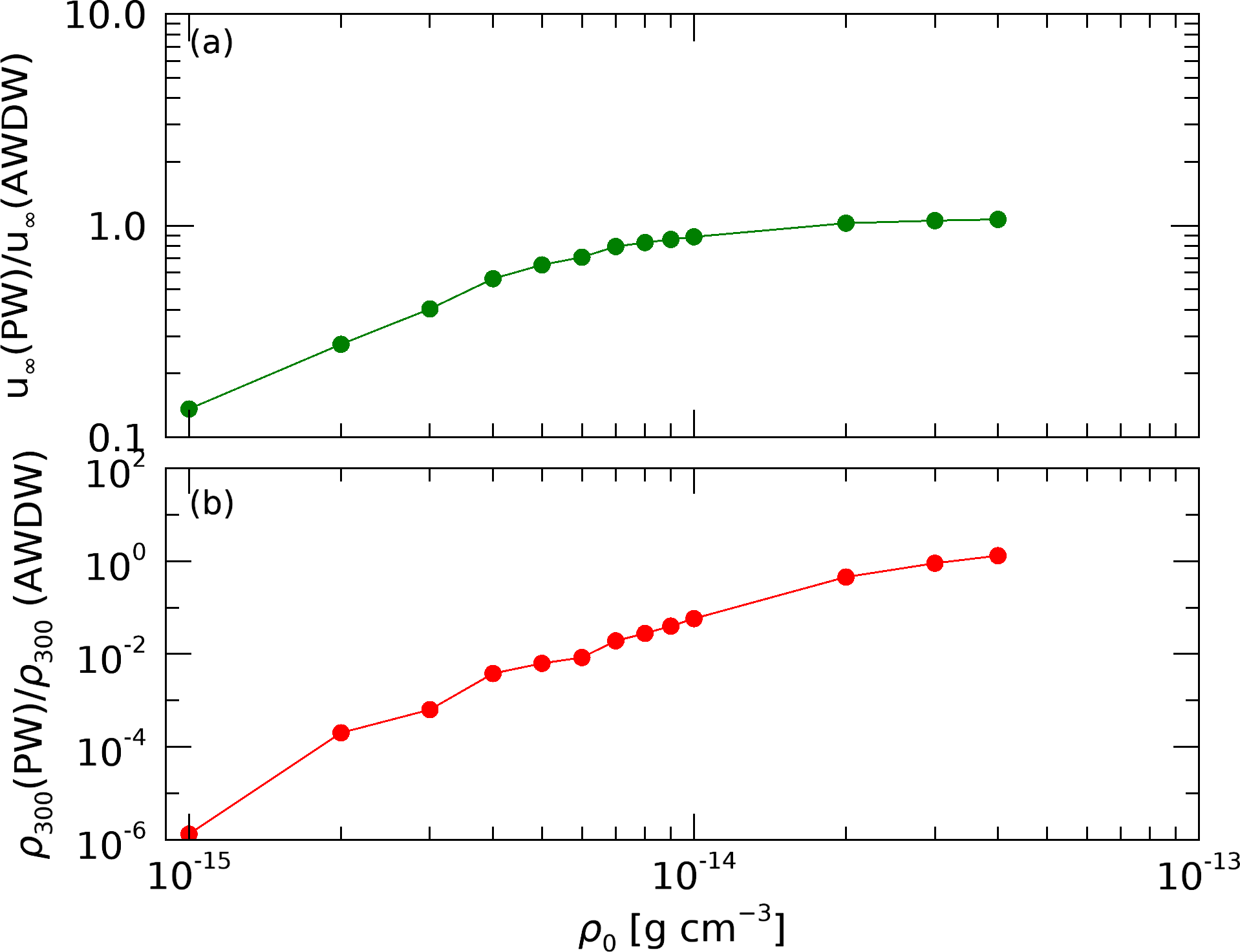}
    \caption{Ratio between the PW and the AWDW as function of base density, for (a) $u_{\infty}$ and (b) $\rho_{300}$. For low density range, the PW shows smaller values for $u_{\infty}$ and underestimated $\rho_{300}$ while for higher densities the two models produce the same result for $u_{\infty}$ and $\rho_{300}$.}
    \label{fig:ratios}
\end{figure}

\Cref{fig:ratios} compares the results from the AWDW and the PW, for $u_{\infty}$ (\Cref{fig:ratios}-a) and $\rho_{300}$ (\Cref{fig:ratios}-b). We see that, for the high base density (high-$\beta$), the ratios are $\sim 1$, showing that both methods reach similar results. In contrast, for the low base density regime (low-$\beta$), $u_{\infty}$ can be nearly one order of magnitude larger for AWDW and $\rho_{300}$ several orders of magnitude larger. Together, these two Figures explain why the PW deviates from the AWDW solution at low-$\beta$ (\Cref{fig:mdot}).

\bsp	
\label{lastpage}
\end{document}